\documentclass{article}

\usepackage{microtype}
\usepackage{graphicx}
\usepackage{subfigure}
\usepackage{booktabs} 

\usepackage{hyperref}

\usepackage[accepted]{icml2025}

\usepackage{amsmath}
\usepackage{amssymb}
\usepackage{mathtools}
\usepackage{amsthm}

\usepackage[capitalize,noabbrev]{cleveref}

\theoremstyle{plain}
\newtheorem{theorem}{Theorem}[section]

\theoremstyle{definition}
\newtheorem{definition}[theorem]{Definition}

\theoremstyle{remark}
\newtheorem{remark}[theorem]{Remark}

\usepackage[textsize=tiny]{todonotes}

\usepackage{caption}
\usepackage{bm}
\usepackage{xcolor}
\usepackage{float}
\usepackage{multirow}
\usepackage{longtable}
\usepackage{framed}
\usepackage{enumitem}
\usepackage{pythonhighlight}
\usepackage{appendix}
\usepackage[utf8]{inputenc}
\usepackage{bbding}
\usepackage{tabularx}
\usepackage{dsfont}
\newcommand{\linecode}[1]{\colorbox[rgb]{1,1,1}{\color{gray} \texttt{#1}}}
\DeclareMathAlphabet{\mathpzc}{OT1}{pzc}{m}{it}

\usepackage{mathtools}
\usepackage{multirow}
\usepackage{algorithm}
\usepackage[noend]{algpseudocode}
\usepackage{color,soul}

\makeatletter
\DeclareRobustCommand\onedot{\futurelet\@let@token\@onedot}
\def\@onedot{\ifx\@let@token.\else.\null\fi\xspace}
\def\eg{\emph{e.g}\onedot} 
\def\ie{\emph{i.e}\onedot} 
 
\def\etc{\emph{etc}\onedot} 
\def\wrt{\emph{w.r.t}\onedot} 

\makeatother

\definecolor{blue_}{RGB}{76, 114, 176}
\definecolor{orange_}{RGB}{221, 132, 82}
\definecolor{upload}{RGB}{47, 85, 151}
\definecolor{download}{RGB}{241, 13, 208}
\definecolor{red_}{RGB}{255, 0, 0}
\definecolor{gray_}{RGB}{127, 127, 127}
\definecolor{green_}{RGB}{1, 128, 0}
\definecolor{sjtured_}{RGB}{192, 0, 0}
\definecolor{sjtugreen_}{RGB}{84, 130, 53}
\definecolor{hist_red}{RGB}{194, 82, 83}
\definecolor{hist_blue}{RGB}{83, 110, 174}

\usepackage{pifont}
\usepackage{makecell}
\usepackage{colortbl}
\definecolor{grayline}{gray}{0.9}
\newcommand{\RomanNumeralCaps}[1]
    {\MakeUppercase{\romannumeral #1}}

\crefname{section}{Sec.}{Secs.}
\Crefname{section}{Section}{Sections}
\Crefname{table}{Table}{Tables}
\crefname{table}{Tab.}{Tabs.}
\crefname{figure}{Fig.}{Figs.}
\crefname{equation}{Eq.}{Eqs.}

\usepackage{wrapfig}
\usepackage{tikz}

\usepackage{xspace}
\def\ours{\texttt{PCEvolve}\xspace}

\icmltitlerunning{\ours: Private Contrastive Evolution for Synthetic Dataset Generation via Few-Shot Private Data and Generative APIs}

\begin{document}

\twocolumn[
\icmltitle{\ours: Private Contrastive Evolution for Synthetic Dataset Generation via Few-Shot Private Data and Generative APIs}



\icmlsetsymbol{equal}{*}

\begin{icmlauthorlist}
\icmlauthor{Jianqing Zhang}{sjtu,thu}
\icmlauthor{Yang Liu}{polyu,ailab}
\icmlauthor{Jie Fu}{sit}
\icmlauthor{Yang Hua}{qub}
\icmlauthor{Tianyuan Zou}{thu}
\icmlauthor{Jian Cao}{sjtu,web}
\icmlauthor{Qiang Yang}{polyu}
\end{icmlauthorlist}

\icmlaffiliation{sjtu}{Shanghai Jiao Tong University}
\icmlaffiliation{polyu}{Hong Kong Polytechnic University}
\icmlaffiliation{sit}{Stevens Institute of Technology}
\icmlaffiliation{qub}{Queen's University Belfast}
\icmlaffiliation{thu}{Institute for AI Industry Research (AIR), Tsinghua University}
\icmlaffiliation{ailab}{Shanghai Artificial Intelligence Laboratory}
\icmlaffiliation{web}{Shanghai Key Laboratory of Trusted Data Circulation and Governance in Web3}

\icmlcorrespondingauthor{Yang Liu}{yang-veronica.liu@polyu.edu.hk}
\icmlcorrespondingauthor{Jian Cao}{cao-jian@sjtu.edu.cn}

\icmlkeywords{Machine Learning, ICML}

\vskip 0.3in
]



\printAffiliationsAndNotice{}  

\begin{abstract}

The rise of generative APIs has fueled interest in privacy-preserving synthetic data generation. While the Private Evolution (PE) algorithm generates Differential Privacy (DP) synthetic images using diffusion model APIs, it struggles with few-shot private data due to the limitations of its DP-protected similarity voting approach. In practice, the few-shot private data challenge is particularly prevalent in specialized domains like healthcare and industry.
To address this challenge, we propose a novel API-assisted algorithm, \textit{Private Contrastive Evolution (\ours)}, which iteratively mines inherent inter-class contrastive relationships in few-shot private data beyond individual data points and seamlessly integrates them into an adapted Exponential Mechanism (EM) to optimize DP's utility in an evolution loop. 
We conduct extensive experiments on four specialized datasets, demonstrating that \ours outperforms PE and other API-assisted baselines. These results highlight the potential of leveraging API access with private data for quality evaluation, enabling the generation of high-quality DP synthetic images and paving the way for more accessible and effective privacy-preserving generative API applications. Our code is available at \url{https://github.com/TsingZ0/PCEvolve}. 

\end{abstract}

\section{Introduction}

Recent advances in machine learning for image tasks have significantly improved efficiency in various fields, such as COVID-19 pneumonia recognition~\cite{harmon2020artificial} and industry anomaly detection~\cite{roth2022towards}. 
However, training an effective model requires a sufficient number of images and computing resources, which is often impractical for resource-constrained clients (\eg, clinics) in specialized domains, where each data producer typically possesses only a few data~\cite{chen2023industrial, chen2020fedhealth}. 
To tackle data scarcity, data owners are increasingly turning to external sources, but most specialized data are private \cite{boland2017ten}, raising significant privacy concerns~\cite{zhang2022privacy}. Healthcare data breaches incur substantial costs, averaging \$9.2 million per year~\cite{hu2024sok}.

\begin{figure}
\centering
\includegraphics[width=\linewidth]{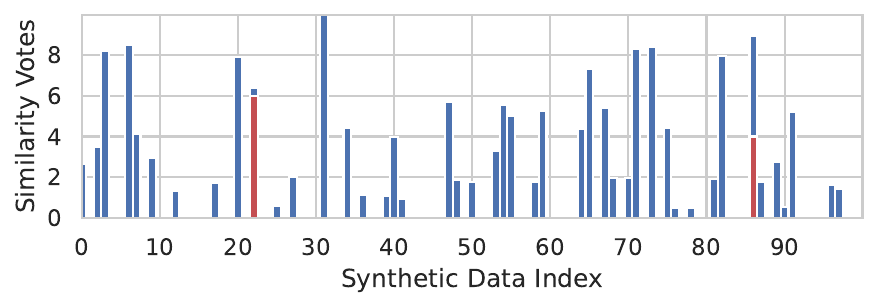}
\caption{A scenario with 10-shot private images and 100-shot synthetic images in PE. Private data contribute only 10 votes (\textcolor{hist_red}{red}), while the noise (\textcolor{hist_blue}{blue}) exceeds the \textcolor{hist_red}{red} votes.}
\label{fig:intro}
\end{figure}

Differential Privacy (DP)~\cite{dwork2006calibrating} is a widely used privacy protection method. However, traditional training-based approaches like DP-GAN~\cite{zhang2024dp} and DP-Diffusion~\cite{ghalebikesabi2023differentially} for image generation are impractical for resource-constrained clients due to their high computational cost and substantial data requirements~\cite{hu2020differentially}. Nowadays, using powerful generative APIs for image generation has emerged as a compelling and training-free alternative~\cite{lindifferentially}. However, since black-box API providers are untrusted, protecting user privacy becomes crucial~\cite{lindifferentially}. 
The latest Private Evolution (PE) algorithm~\cite{lindifferentially} leverages diffusion model APIs~\cite{rombach2022high} to generate DP synthetic images without any training in an evolution loop. Specifically, PE generates random synthetic images from APIs and iteratively improves them by selecting (with DP) the most similar synthetic images to the private dataset, and then querying APIs to generate more similar images. DP synthetic data can be reused infinitely for potential models on various downstream tasks without additional privacy costs, thanks to the post-processing property of DP~\cite{dwork2006calibrating, hu2024sok, fu2024dpsur, fu2024differentially}. 

However, PE was originally designed for data parties with large-scale private image datasets, and its performance degrades with limited private data due to its reliance on DP-protected (based on Gaussian Mechanism (GM)~\cite{dwork2014algorithmic}) similarity voting, where one private data gives one vote to the most similar synthetic data with DP. For example, the original PE uses 302,436 images from the Camelyon17 dataset~\cite{koh2021wilds} as private data\footnote{PE also considers a Cat dataset with 200 images, but only shows synthetic images with a large domain gap to private data.}. 
However, as shown in \cref{fig:intro}, when dealing with few-shot private data, the added noise overwhelms the actual votes, leading to nearly random similarity voting and selection. 
Subsequently, querying APIs with randomly selected images as guidance yields nearly random outputs, ultimately degrading the quality of synthetic data~\cite{pan2023arbitrary}. 

To address the few-shot private data challenge in API-assisted DP generation inside the evolution loop, we propose an algorithm called \textit{Private Contrastive Evolution (\ours)} by incorporating two key components for selecting high-quality prototypical synthetic data as feedback to APIs:

(1) We devise a \textit{contrastive filter} to iteratively exploit inter-class contrastive relationships between different classes within the private data, beyond individual data points, thereby enhancing the class-discriminability of synthetic data \wrt private data. 



(2) We adapt the Exponential Mechanism (EM)~\cite{dwork2014algorithmic} to preserve private inter-class contrastive relationships, addressing the excessive noise from the high sensitivity of GM. Specifically, we devise a \textit{similarity calibrator} to enhance EM's utility by prioritizing high-quality synthetic data that closely resembles private data.


To evaluate the effectiveness of \ours, we conduct extensive experiments in two aspects:

(1) We show that \ours surpasses six API-assisted synthetic data generation baselines across four practical datasets from specialized domains like healthcare and industry. Additionally, we demonstrate the applicability of \ours in various scenarios, such as varying amounts of private/synthetic data, different APIs, etc. 
Our DP synthetic dataset works effectively with six diverse downstream models, thanks to its enhanced class-discriminability. 

(2) We analyze the properties of \ours, including the visible quality of synthetic images, the effectiveness of each component, and the influence of the single hyperparameter. 

\section{Related Work}


\textbf{Synthetic image data with APIs.} 
In recent years, numerous studies have focused on synthetic image generation using generative models, such as Stable Diffusion (SD)~\cite{rombach2022high, moor2023foundation, yang2023diffusion, zhang2023adding, brooks2023instructpix2pix}.
However, in specialized domains like healthcare, generative models often require fine-tuning or even redesigning on local platforms with GPU resources~\cite{ji2024diffusion, guan2021domain, kather2022medical}, demanding significant human effort and financial investment—making them impractical for resource-constrained clients~\cite{abou2024parameter, wornow2023shaky}. 
An alternative is leveraging cloud-based APIs from service providers~\cite{he2023synthetic, seo2024just, samuel2024generating}, customizing them through API-based fine-tuning~\cite{openai_finetuning_guide} or prompt engineering~\cite{hao2024optimizing}. 
While this approach reduces resource consumption, it raises serious privacy concerns~\cite{qifine}, as local private data must be uploaded to untrusted API providers for fine-tuning or prompting~\cite{chen2024janus}. 
Although some API-assisted text data generation methods~\cite{ye2022progen, gaoself} avoid using private data, they rely on specific downstream models, limiting their utility. 

\textbf{DP synthetic image data with APIs.}
DP~\cite{dwork2006calibrating} is a widely used technique for protecting privacy in image data~\cite{ziller2021medical} and has also been applied to synthetic image generation~\cite{li2024privimage, hu2024sok, chen2022dpgen, de2024synthetic}. 
However, DP-based generative methods like DP-GAN~\cite{zhang2024dp} and DP-Diffusion~\cite{ghalebikesabi2023differentially} are impractical for resource-constrained clients~\cite{hu2020differentially} and infeasible when models are accessed via APIs~\cite{xiedifferentially, lindifferentially}. 
Directly adding DP noise to images makes DP images nearly unusable as input guidance for APIs, as ensuring privacy demands excessive noise~\cite{croft2021obfuscation}. 
Recently, Private Evolution (PE)~\cite{lindifferentially} optimizes synthetic image data to resemble private data with DP, performing well with large private datasets but \textit{struggling in few-shot scenarios due to its GM-based similarity voting}. Two follow-up methods in the \textit{text} domain~\cite{xiedifferentially, houpre} retain PE's core GM-based similarity voting while equipping it with more synthetic data per generation~\cite{xiedifferentially} or more private data from large-scale federated learning~\cite{houpre}. 

\section{Preliminaries}

\begin{definition}[Differential Privacy~\cite{dwork2006calibrating}]
    Let $\mathcal{D}$ be any private database with data from a space $\mathbb{X}$, denoted as $\mathcal{D} \in \mathbb{X}$, with a symmetric neighbor $\mathcal{D}'$, where $\mathcal{D}$ and $\mathcal{D}'$ differ in only one element. Let $\epsilon > 0$ and $\delta \in [0,1]$ be two privacy parameters. A randomized algorithm $\mathcal{M}: \mathbb{X} \rightarrow \mathbb{Y}$ with a value range $\mathbb{Y}$ is $(\epsilon, \delta)$-DP if the following inequality holds for any $\mathcal{E} \subseteq \mathbb{Y}$: 
    $$
    Pr[\mathcal{M}(\mathcal{D})\in \mathcal{E}] \le e^\epsilon Pr[\mathcal{M}(\mathcal{D}')\in \mathcal{E}] + \delta. 
    $$
    If $\delta=0$, we say that $\mathcal{M}$ is $\epsilon$-DP. Any post-processing of a DP mechanism's output incurs no additional privacy loss. 
\end{definition}

\begin{definition}[Gaussian Mechanism~\cite{dwork2014algorithmic}]
    Let $f: \mathbb{X} \rightarrow \mathbb{R}^D$ be a $D$-dimensional function with $\ell_2$ sensitivity to be $\Delta_f := \max_{\mathcal{D}, \mathcal{D}'} ||f(\mathcal{D}) - f(\mathcal{D}')||_2$. The Gaussian Mechanism (GM) $\mathcal{M}_{\sigma}$ with parameter $\sigma$ adds noise scaled to $\mathcal{N}(0, \sigma^2)$ to each of the $D$ components of the output, \ie, 
    $
    \tilde{f}(\mathcal{D}) := f(\mathcal{D}) + \mathcal{N}(\bf{0}, \sigma^2 \bf{I}_D). \label{def:fD}
    $
    For $\epsilon, \delta \in (0, 1)$, $\mathcal{M}_{\sigma}$ with $\sigma = \Delta_f\sqrt{2\log(1.25/\delta)}/\epsilon$ is $(\epsilon, \delta)$-DP.
\end{definition}

\begin{definition}[Exponential Mechanism~\cite{dong2020optimal, mcsherry2007mechanism}]
    Given a parameter $\epsilon$, an arbitrary range $\mathcal{R}$, and a utility function $u: \mathbb{X}\times \mathcal{R} \rightarrow \mathbb{R}$ with sensitivity $\Delta_u := \max_{r\in \mathcal{R}} \max_{\mathcal{D}, \mathcal{D}'} |u(\mathcal{D}, r) - u(\mathcal{D}', r)|$, a randomized algorithm $\mathcal{M}_u$ is called the Exponential Mechanism (EM), if the outcome $r$ is sampled with probability proportional to $\exp{(\frac{\epsilon \cdot u(\mathcal{D}, r)}{2\Delta_u})}$ and $\mathcal{M}_u$ is $\epsilon$-DP: 
    $$
    Pr[\mathcal{M}_u(\mathcal{D})=r] = \frac{\exp{(\frac{\epsilon \cdot u(\mathcal{D}, r)}{2\Delta_u})}}{\sum_{r'\in \mathcal{R}}\exp{(\frac{\epsilon \cdot u(\mathcal{D}, r')}{2\Delta_u})}}. \label{def:exp}
    $$
\end{definition}

\begin{definition}[Sequential Composition~\cite{dwork2006calibrating}]\label{def:Sequential}
    Given any mechanism $\mathcal{M}_{1}(\cdot)$  that satisfies  $\epsilon_{1}$-DP, and  $\mathcal{M}_{2}(s, \cdot)$  that satisfies  $\epsilon_{2}$-DP for any $s$, then  $\mathcal{M}(\mathcal{D})=\mathcal{M}_{2}\left(\mathcal{M}_{1}(\mathcal{D}), \mathcal{D}\right)$  satisfies  $\left(\epsilon_{1}+\epsilon_{2}\right)$-DP . 
\end{definition}

\begin{remark}
    Given the same $\epsilon$, $\mathcal{M}_u$ ($\epsilon$-DP) provides stronger privacy protection than $\mathcal{M}_{\sigma}$ ($(\epsilon, \delta)$-DP) for $\delta > 0$ .
\end{remark}

\section{Method}


\textbf{Motivation.} The Gaussian DP ( $\mathcal{M}_{\sigma}$)-based similarity voting approach in the existing method PE corresponds to the function $f$ (in \cref{def:fD}) of $\mathcal{M}_{\sigma}$. However, due to the few-shot private dataset $\mathcal{D}_p$, the values in $f(\mathcal{D}_p)$ are extremely small, making the noise from $\mathcal{N}(0, \sigma^2)$ appear larger than $f(\mathcal{D}_p)$ and rendering the final votes, $\tilde{f}(\mathcal{D}_p)$, nearly random, as shown in \cref{fig:intro}. 
In contrast, EM $\mathcal{M}_u$ is tailored for selection, with privacy guarantee depending on the sensitivity of utility function $u$~\cite{dwork2008differential}. 
Therefore, we propose \ours based on $\mathcal{M}_u$ and adapt EM to select synthetic (public) data instead of private data, reducing the influence of private data volume and making it suitable for few-shot scenarios. 

\textbf{Problem Definition.}
Our goal is to generate a DP synthetic dataset $\mathcal{D}_s$ that closely resembles $\mathcal{D}_p$ and effectively supports potential downstream $C$-class classification tasks within the same domain as $\mathcal{D}_p$. To achieve this, $\mathcal{D}_s$ must satisfy two key criteria: (1) \textit{class-discriminability} and (2) high \textit{similarity} to $\mathcal{D}_p$. 

Following PE, we randomly initialize $\mathcal{D}^0_s$ using an untrusted text-to-image (t2i) API $G_{t2i}$ with a \textit{simple} text prompt $\mathcal{T}$—containing only the domain and class label names, \textit{without any prompt engineering}~\cite{mesko2023prompt} (\eg, $\mathcal{T} =$ ``A leather texture image with cut defect''). We then iteratively refine $\mathcal{D}_s$ (omitting the iteration superscript $t$ for simplicity) through the following steps.
In each iteration, given the previously generated $\mathcal{D}_s$ and a pre-trained encoder\footnote{All distance measures are computed in the feature space after applying $E_f$. For simplicity, we omit $E_f$ in the following.} (feature extractor) $E_f$, we select high-quality prototypical (``proto'') data points, denoted as $\mathcal{D}_{pro}$, from $\mathcal{D}_s$. This selection is guided by $\mathcal{D}_p$ using $\mathcal{M}_u$.
Next, we refine the synthetic dataset by leveraging an untrusted image-to-image (i2i) API $G_{i2i}$, using $\mathcal{D}_{pro}$ as feedback and guidance. As this evolution loop progresses, we iteratively optimize $\mathcal{D}_s$.

A crucial aspect of this process is the selection of $\mathcal{D}_{pro}$. According to \cref{def:exp} of $\mathcal{M}_u$, $\mathcal{D}_{pro}$ is obtained by sampling data from $\mathcal{D}_s$ with probabilities proportional to their corresponding $u$ scores. Thus, the effectiveness of our approach hinges on the design of the utility function $u$.

\textbf{Threat Model.}
We assume that the API provider is honest-but-curious, aiming to extract private information (\eg, membership information) from the private data uploaded by the client device. While the provider can only access public images selected by the client’s private data, it can leverage this information to attack the client’s privacy~\cite{duan2024flocks}. Specifically, by repeatedly observing the output images, the provider can infer whether a particular sample exists in the client’s dataset. Empirical studies have confirmed privacy risks through viable attacks, such as membership inference attacks~\cite{shokri2017membership,carlini2022membership}.


\subsection{\ours}

\begin{figure*}[h]
\centering
\includegraphics[width=\linewidth]{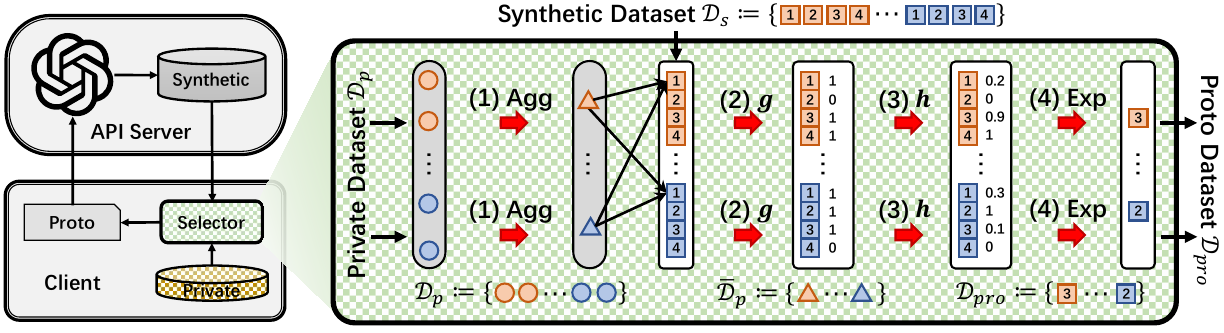}
\caption{Illustration of our \ours, whose core is the DP-protected selector. Different colors denote distinct data classes (two classes are explicitly shown, with others indicated by ``$\ldots$''). ``Agg'' and ``Exp'' denote the class center aggregation and the EM \(\mathcal{M}_u\) ($u = h \circ g$), respectively.}
\label{fig:ours}
\end{figure*}


\textbf{Overview.} 
As shown in \cref{fig:ours}, our \ours iteratively enhances the synthetic data with a DP-protected selector as the core engine. To tackle the few-shot challenge with DP, we leverage inter-class contrastive relationships from private data and optimize the utility of $\mathcal{M}_u$ within the selector. 
Specifically, we (1) aggregate private class centers to reduce bias caused by the few-shot problem, (2) introduce a \textit{contrastive filter} $g$ to improve the class-discriminability of synthetic data, (3) design a \textit{similarity calibrator} $h$ to maximize the selection probability of the most similar synthetic data by $\mathcal{M}_u$, where $u := h \circ g$, and (4) utilize the scores from $u$ (\ie, outputs of $h$) to construct $\mathcal{D}_{pro}$ via $\mathcal{M}_u$.

\textbf{(1) Aggregating Class Centers}. The fundamental challenge in few-shot data scenarios is the lack of sufficient information to represent the entire real data distribution, leading to inadequate and biased learning of downstream models~\cite{song2023comprehensive, xu2022alleviating}, especially the few but useful data that are near the distribution boundary~\cite{yangfree}. To reduce bias from boundary data, we aggregate few-shot private data to obtain the private center set $\bar{\mathcal{D}}_p:= \{\bar{d}^c_p\}_{c \in [C]}$, where $\bar{d}^c_p := \frac{1}{|\mathcal{D}^c_p|} \sum_{d^c_p \sim \mathcal{D}^c_p} d^c_p$, where we define $\mathcal{D}^c$ as the subset of any set $\mathcal{D}$ that contains all data with label $c$. 

\textbf{(2) Contrastive Filter $g$}. Then, we leverage inter-class contrastive relationships to utilize private data fully. Inspired by metric learning~\cite{kulis2013metric} and contrastive learning~\cite{chen2020simple}, we design a contrastive filter $g$ to select synthetic data that can be correctly classified into their corresponding classes using $\bar{\mathcal{D}}_p$ as class identifiers, \ie, 
\begin{equation}
    g(d_s^c, \bar{\mathcal{D}}_p) :=
        \begin{cases} 
        1, & \text{if } \ell_2(d_s^c, \bar{d}_p^c) < \min_{c'} \{\ell_2(d_s^c, \bar{d}_p^{c'})\} \\
        0, & \text{otherwise}
        \end{cases}
        , \label{eq:g}
\end{equation}
$\forall d^c_s \in \mathcal{D}^c_s, c\in [C], c' \in [C], c' \neq c$. We use $\ell_2$-norm~\cite{luo2016regression} for distance measure.
Using $g$ for selection aligns our discriminability goal by assigning positive scores to discriminative synthetic data. 

\textbf{(3) Similarity Calibrator $h$.} Given the initial large difference between the synthetic and private domains, solely emphasizing discriminability with $g$ while neglecting similarity will barely reduce this gap in the next generation iteration. Thus, we calculate the similarity of discriminative synthetic data to $\bar{\mathcal{D}}_p$ by
\begin{equation}
    h(d_s^c, \bar{\mathcal{D}}_p) :=
        \begin{cases} 
        e^{-\ell_2(d_s^c, \bar{\mathcal{D}}_p)}, & \text{if } g(d_s^c, \bar{\mathcal{D}}_p) = 1 \\
        0, & \text{otherwise}
        \end{cases}
        , \label{eq:h1}
\end{equation}
$\forall d^c_s \in \mathcal{D}^c_s, c\in [C]$. We apply exponent arithmetic and negation to convert $\ell_2$ distances into similarities. Since the range of $\ell_2$ is $[0, +\infty)$, the range of $h$ is $[0, 1]$, making $u = h \circ g$ with a sensitivity of $\Delta_u = 1$ according to \cref{def:exp}. Since $\mathcal{M}_u$ samples data from $\mathcal{D}_s$ into $\mathcal{D}_{pro}$ based on $u$'s values, candidates with higher $u$ values are more likely to be included in $\mathcal{D}_{pro}$. 

In practice, $\ell_2$ values rarely span the full range of $[0, +\infty)$. Due to random initialization and the large gap between $\mathcal{D}_s$ and $\mathcal{D}_p$, initial $\ell_2$ values are excessive, leading most $u$ values to fall near $0$. 
This under-utilizes the full range $[0,1]$ of $u$ and 
causes $\mathcal{M}_u$ to sample data almost randomly. 
To address this issue, we propose calibrating the original similarity scores to ensure $u$ values span the full range of $[0,1]$ in two steps: (1) normalizing the original $\ell_2$ values to $[0,1]$ and (2) scaling them to $[0,\tau]$, where $\tau$ is the only hyperparameter. This forces $u$ values to fall within $\{0\} \cup [e^{-\tau},1]$. 
By selecting an appropriate $\tau$ value, we can ensure that $e^{-\tau} \approx 0$.
Consequently, we assign the maximum $u$ value (\ie, $1$) to the best (the most similar) candidate to maximize its selection probability when applying $\mathcal{M}_u$. Formally, we rewrite $h$ to 
\begin{equation}
\begin{aligned}
    h(d_s^c, \bar{\mathcal{D}}_p) &:=
    \begin{cases}
    e^{-\frac{\ell_2(d_s^c, \bar{\mathcal{D}}_p) - \ell^c_{\min}}{\ell^c_{\max} - \ell^c_{\min}} \cdot \tau}, & \text{if } g(d_s^c, \bar{\mathcal{D}}_p) = 1 \\
    0, & \text{otherwise}
    \end{cases},
    \\
    \text{s.t. } \ &\ell^c_{\max} := \max_{d_s^c \in \mathcal{D}_s^c, \ g(d_s^c, \mathcal{D}_p) = 1} \ell_2(d_s^c, \bar{\mathcal{D}}_p), \\
    &\ell^c_{\min} := \min_{d_s^c \in \mathcal{D}_s^c, \ g(d_s^c, \mathcal{D}_p) = 1} \ell_2(d_s^c, \bar{\mathcal{D}}_p), \label{eq:h2}   
\end{aligned}
\end{equation}
$\forall d^c_s \in \mathcal{D}^c_s, c\in [C]$. As per \cref{def:exp}, this rewriting does not impact $\Delta_u$, since $u$ values still belong to $[0, 1]$. 

\textbf{(4) Applying $\mathcal{M}_u$.} Given $u$ and $\Delta_u$, we simply run $\mathcal{M}_u$ to sample synthetic data from $\mathcal{D}_s$ into $\mathcal{D}_{pro}$ based on \cref{def:exp}. Each execution of $\mathcal{M}_u$ consumes a portion of the total privacy cost $\epsilon_*$. To balance the utility-privacy trade-off~\cite{dwork2014algorithmic}, we select only one candidate per class to form $\mathcal{D}_{pro}$. The best candidate with the highest $h$ value has the greatest probability of being selected. 
Then, we can generate a refined $\mathcal{D}_s$ using an improved $\mathcal{D}_{pro}$ as the input to $G_{i2i}$. We show the overall algorithm in \cref{algo}.
\begin{algorithm}[ht]
\caption{\ours}
\begin{algorithmic}[1]
    \Require Private dataset $\mathcal{D}_p$,  i2i API $G_{i2i}$, t2i API $G_{t2i}$, text prompt $\mathcal{T}$, total privacy cost $\epsilon_*$, number of class $C$, number of iteration $T$, encoder $E_f$, and similarity calibrating factor $\tau$. 
    \Ensure Synthetic dataset $\mathcal{D}_s$.
    \State $\mathcal{D}^0_s \gets G_{t2i}(\mathcal{T})$ and $\epsilon = \frac{\epsilon_*}{T\cdot C}$.
    \For{evolution iteration $t=1, \ldots, T$}
        \For{class $c=1, \ldots, C$}
            \State Get $u$ scores for $\mathcal{D}^{t,c}_s$ via \cref{eq:g} and \cref{eq:h2}.
            \State Get $\mathcal{D}^{t,c}_{pro}$ by sampling data from $\mathcal{D}^{t,c}_s$, with 
            \Statex \qquad \quad the index set $\mathcal{R}$ of $\mathcal{D}^{t,c}_s$ and probabilities
            $$
            Pr[\mathcal{M}_u(\mathcal{D}_p)=r\in \mathcal{R}] = \frac{\exp{(\frac{\epsilon \cdot u(\mathcal{D}_p, r)}{2\Delta_u})}}{\sum_{r'\in \mathcal{R}}\exp{(\frac{\epsilon \cdot u(\mathcal{D}_p, r')}{2\Delta_u})}}.
            $$
        \EndFor
        \State $\mathcal{D}^{t}_s \gets G_{i2i}(\mathcal{T}, \mathcal{D}^{t}_{pro})$, where $\mathcal{D}^{t}_{pro} = \{\mathcal{D}^{t,c}_{pro}\}^{C}_{c=1}$.
    \EndFor
    \\
    \Return Synthetic dataset $\mathcal{D}^T_s$.
\end{algorithmic}
\label{algo}
\end{algorithm}


\subsection{Privacy Analysis}

\begin{theorem}\label{the:dp} \cref{algo} \ours satisfied $\epsilon_*$-DP.
\end{theorem}
\begin{proof}
Firstly, we conduct a privacy analysis for each query to privacy dataset $\mathcal{D}_p$ (Line 5 in \cref{algo}). For any synthetic data set $\mathcal{D}_s$ and its corresponding indexes set $\mathcal{R}$, the sensitivity $\Delta_u$ between two adjacent datasets $\mathcal{D}_p,\mathcal{D}_p^{\prime}\in \mathcal{D}$ is bound by 1 according to our scoring function $u = h \circ g$:
$$
\Delta_u=\max_{r\in\mathcal{R}}\max_{\mathcal{D}_p,\mathcal{D}_p^{\prime}\in\mathcal{D}}|u(\mathcal{D}_p,r)-u(\mathcal{D}_p^{\prime},r)|=1.\label{equ:a-alpha}
$$
Thus, based on \cref{def:exp}, each time privacy dataset $\mathcal{D}_p$ satisfies $\epsilon$-DP.

Since we access $T\times C$ times to private data $\mathcal{D}_p$, according to the \cref{def:Sequential}, finally our \ours satisfies $\epsilon_*$-DP.
\end{proof}

\section{Experiments}

\subsection{Setup}
\label{sec:setup}

\textbf{Image Generation APIs.} We consider three image generation APIs: Stable Diffusion (SD)~\cite{rombach2022high}, SD with the IP-Adapter~\cite{ye2023ip} plug-in (SD+IPA), and online OpenJourney~\cite{openjourney} API (OJ (online)). We primarily use the widely adopted SD API~\cite{lindifferentially} as the image generation API. Following PE, we generate $N$-shot synthetic images per class in the dataset $\mathcal{D}_s$, with a default setting of $N=100$. In our scenario, an excessively large $N$ is impractical for a resource-constrained client to generate image data using generative APIs. 

\textbf{Few-Shot Datasets.} 
We evaluate \ours on four datasets across two specialized domains under $K$-shot settings.
In healthcare, we use (1) COVIDx~\cite{wang2020covid} (chest X-ray images for COVID-19, two classes), (2) Camelyon17~\cite{koh2021wilds} (tumor tissue patches from breast cancer metastases, two classes), and (3) KVASIR-f (endoscopic images for gastrointestinal abnormal \underline{f}indings detection subset from KVASIR~\cite{pogorelov2017kvasir}, three classes). In industry, we use MVTecAD-l (\underline{l}eather surface anomaly detection subset from MVTecAD~\cite{bergmann2019mvtec}, three classes). By default, we set $K=10$, as MVTecAD-l has only 19 images per class. This value of $K$ is typical for few-shot image tasks~\cite{he2023synthetic}.

\textbf{Baselines.} We compare \ours with six baselines across three categories, all of which focus on generating image datasets using untrusted black-box API(s), without training: 

(\RomanNumeralCaps{1}) \textit{Using t2i APIs for image generation}: \\
$\bullet$~B~\cite{he2023synthetic}, which uses only a t2i API with a simple text prompt $\mathcal{T}$ that includes only the domain and class label name. 
$\bullet$~LE~\cite{seo2024just}, which extends B with a LLaMA~\cite{touvron2023llama} API to enhance $\mathcal{T}$. 
$\bullet$~RF~\cite{samuel2024generating}, which filters out bad t2i-generated data that closely resemble private data from different classes. 
$\bullet$~GCap, which generates images using a t2i API with a LLaVA~\cite{liu2024visual} API for extracting private image captions. 

(\RomanNumeralCaps{2}) \textit{Using i2i APIs for image generation with DP}: \\
$\bullet$~DPImg, which directly adds DP (GM) noise to few-shot private images to generate DP replicas, which are then input to an i2i API. DPImg adapts RG~\cite{he2023synthetic} to ensure DP while avoiding modifications to the generative API.

(\RomanNumeralCaps{3}) \textit{Using t2i and i2i APIs for image generation with DP}: \\
$\bullet$~PE~\cite{lindifferentially}, like \ours, generates DP synthetic image datasets using private data along with both t2i and i2i APIs within a privacy-preserving evolution loop. 

\textbf{Implementation Details.} To maximize performance in few-shot scenarios following~\citet{he2023synthetic}, we train a new classification head for a given pre-trained backbone model on the final synthetic dataset $\mathcal{D}_s$. We report the Top-1 accuracy on the entire downstream test sets from the above datasets\footnote{For simplicity, we focus on balanced sets, following PE.}. Top-1 accuracy, also known as the classification accuracy score (CAS)~\cite{ravuri2019classification}, is a widely used metric for assessing the quality of synthetic datasets in downstream tasks~\cite{frolov2021adversarial, lee2024holistic}. By default, we use ResNet-18~\cite{he2016deep} as the pre-trained backbone model and encoder $E_f$ due to its broad applicability across resource-constrained clients. 
For DP methods, the overall privacy cost $\epsilon_*$ is set to 10, 8, 8, and 10 for COVIDx, Camelyon17, KVASIR-f, and MVTecAD-l, respectively. For GM in DPImg and PE, we set $\delta$ to $10^{-5}$, higher than EM's $0$ in \ours. 
We run each experiment three times and report the mean. Please refer to the specific experiments and Appendix for more details and results.

\subsection{Performance of \ours}

\textbf{CAS \wrt Four Specialized Datasets}

\begin{table}[h]
\centering
\caption{Top-1 accuracy (\%) on four specialized datasets. }
\resizebox{\linewidth}{!}{
\begin{tabular}{l|*{4}{c}}
\toprule
 & COVIDx & Came17 & KVASIR-f & MVAD-l \\
\midrule
\textcolor{gray_}{Init} & \textcolor{gray_}{49.34} & \textcolor{gray_}{50.47} & \textcolor{gray_}{33.43} & \textcolor{gray_}{33.33} \\
\midrule
RF & 50.01 & 54.82 & 34.66 & 48.17 \\
GCap & 50.86 & 55.77 & 32.66 & 27.33 \\
B & 50.42 & 54.41 & 32.57 & 43.21 \\
LE & 50.02 & 55.44 & 35.51 & 27.93 \\
\midrule
DPImg & 49.14 & 61.06 & 33.35 & 37.03 \\
\midrule
PE & 59.63 & 63.66 & 48.88 & 57.41 \\
PE-EM & 57.60 & 63.34 & 43.01 & 50.06 \\
\ours-GM & 56.91 & 62.63 & 43.55 & 55.56 \\
\rowcolor{grayline}
\ours & \textbf{64.04} & \textbf{69.10} & \textbf{50.95} & \textbf{59.26} \\
\bottomrule
\end{tabular}}
\label{tab:dataset}
\end{table}

We first assess \ours across four datasets from specialized domains like healthcare and industry in \cref{tab:dataset}, where ``Init'' refers to the downstream models with initial heads. ``Came17'' and ``MVAD'' abbreviate Camelyon17 and MVTecAD for space efficiency, respectively. For category \RomanNumeralCaps{3}, we additionally consider two variants of PE and \ours as baselines: PE-EM, which applies the EM $\mathcal{M}_u$ to PE’s original similarity votes, and \ours-GM, which applies the GM $\mathcal{M}_{\sigma}$ to \ours’s $u$ values. 

As shown in \cref{tab:dataset}, the methods from category \RomanNumeralCaps{3} outperform the others, with \ours achieving the best performance among them. 
Specifically, \ours surpasses baselines by up to 5.44\% in accuracy on Came17. 
This phenomenon stems from the privacy-preserving evolution loop, which iteratively improves image quality while maintaining privacy within a given cost (\ie, the privacy cost $\epsilon_*$). 
RF and GCap indirectly utilize private images via private-image-based post-filtering and captions. Without privacy-preserving techniques, they leak more privacy than DP-protected methods~\cite{sanderdifferentially}, despite accessing private data only once without evolution. In contrast, B and LE use only text prompts, ensuring full privacy protection. However, their performance is slightly below RF and GCap. 
DPImg adds DP noise to private images, rendering them unrecognizable for required privacy, leading to poor or negative performance~\cite{croft2021obfuscation}, especially on hard tasks, \eg, COVIDx and KVASIR-f. 
All methods perform better in less specialized domains (roughly, for two-class datasets, COVIDx (chest X-ray) is harder than Came17 (tumor tissue), and for three-class datasets, KVASIR-f (medical) is harder than MVAD-l (industrial)). Additionally, APIs (SD, LLaMA, LLaVA, \etc.) do not always enhance performance, as they can introduce noise~\cite{barman2024dark}.

Within category \RomanNumeralCaps{3}, PE-EM and \ours-GM underperform PE and \ours. PE-EM lacks similarity score calibration, weakening the EM, while \ours-GM applies the GM to $u$ scores, requiring high sensitivity ($\Delta_f = K \times N$) and reducing utility. By design, the similarity votes in PE align with the GM, while the calibrated $u$ scores in \ours match the EM. Compared to PE, \ours is better suited for specialized domains with few-shot private data, as we incorporate inter-class contrastive relationships and maximize the EM’s effectiveness under the same privacy cost. 

\textbf{CAS \wrt $K$-Shot Private Data}

\begin{figure}[ht]
\centering
\includegraphics[width=\linewidth]{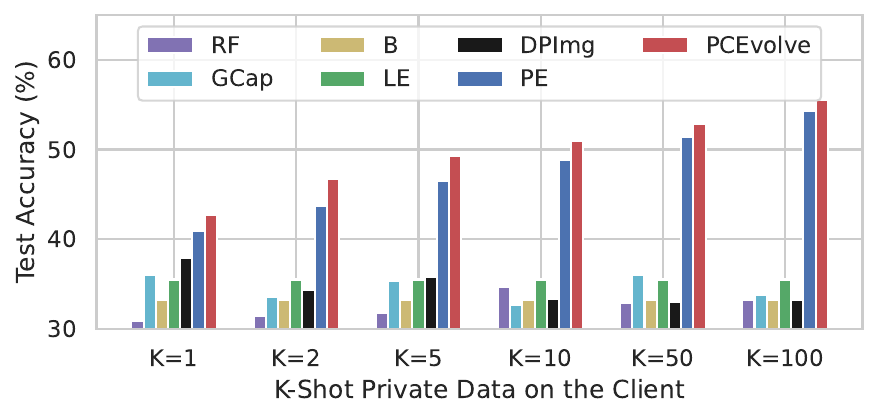}
\caption{Top-1 accuracy of ResNet-18 on KVASIR-f with varying shots of private data per class.}
\label{fig:k}
\end{figure}

In the previous experiments, we used the default setting of $K=10$ for $K$-shot private data on the client. We now explore different values of $K$ to examine how varying the amount of private data affects downstream models. As shown in \cref{fig:k}, methods that leverage private data to filter or evaluate generated synthetic data, such as RF, PE, and \ours, tend to perform better given more private data. The performance improvement is particularly noticeable for PE and \ours, since both methods rely on an evolution loop that iteratively accesses private data multiple times. This repeated access allows them to benefit more from richer private data, leading to a greater enhancement in the quality of the synthetic data generated and, consequently, the downstream model. 
In scenarios with extremely small amounts of private data, such as $K=1$, the class center aggregation subroutine in our \ours becomes invalid, as the aggregated private class centers are identical to the private data. Despite this, \ours still outperforms other methods with our two key components: contrastive filter ($g$) and similarity calibrator ($h$). 
The performance of B and LE remains unaffected by $K$ since they do not require private data. 
DPImg performs worse as the amount of private data increases since DP requires adding more noise to ensure privacy. 

\textbf{CAS \wrt $N$-Shot Synthetic Data}

\begin{figure}[h]
\centering
\includegraphics[width=\linewidth]{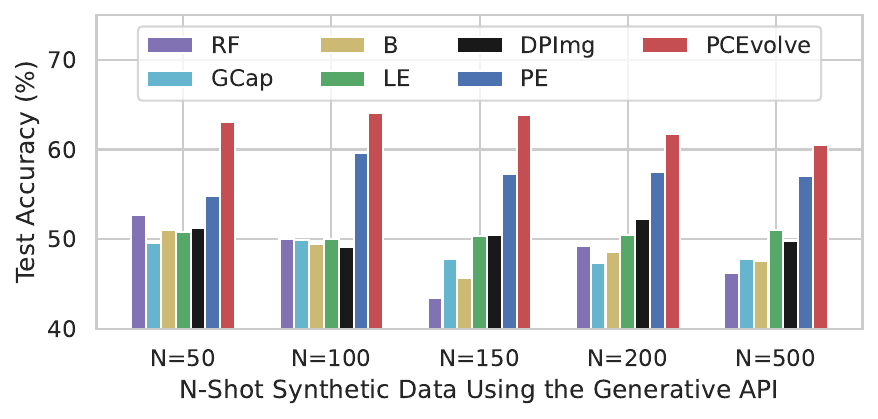}
\caption{Top-1 accuracy of ResNet-18 on COVIDx with varying synthetic data shots per class per iteration.}
\label{fig:n}
\end{figure}

We default to generating $N=100$ synthetic images per class on the client. This choice helps mitigate the risk of generative APIs introducing noise due to insufficient contextual information~\cite{ronanki2024prompt, wang2023reprompt}, which, in our scenario, comes from $K$-shot ($K=10$) private data. 
The noise becomes even more pronounced when generating a large volume of synthetic data, as shown in~\cref{fig:n} ($N=150/200/500$), where accumulated noise and potential inaccuracies degrade the overall quality and utility of the dataset. Moreover, a larger $N$ increases API resource consumption, which may become unaffordable for resource-constrained clients. On the other hand, if the synthetic data amount is too small (\eg, $N=50$), there is insufficient knowledge for downstream models to effectively learn from the generative APIs. 
However, our \ours maintains its performance even at $N=50$, with only a $0.95\%$ drop in accuracy, whereas PE suffers a $4.78\%$ decline. This demonstrates that \ours is more adaptable to resource-constrained clients, highlighting its practical value.

\textbf{CAS \wrt Various APIs}

\begin{table}[h]
\centering
\caption{Top-1 accuracy (\%) on COVIDx and KVASIR-f using SD+IPA and OJ (online) APIs.}
\resizebox{\linewidth}{!}{
\begin{tabular}{l|cc|cc}
\toprule
 & \multicolumn{2}{c|}{COVIDx}  & \multicolumn{2}{c}{KVASIR-f}  \\
\midrule
APIs & SD+IPA & OJ (online) & SD+IPA & OJ (online) \\
\midrule
RF    & 45.03 & 47.91 & 27.22 & 36.55 \\
GCap    & 53.70 & 47.42 & 28.77 & 37.11 \\
B     & 46.61 & 50.22 & 26.27 & 36.61 \\
LE    & 49.79 & 53.17 & 31.22 & 37.38 \\
\midrule
DPImg & 50.58 & 49.61 & 36.89 & 35.05 \\
\midrule
PE    & 56.92 & 54.47 & 48.83 & 48.17 \\
\rowcolor{grayline}
\ours & \textbf{60.46} & \textbf{65.88} & \textbf{52.77} & \textbf{54.58} \\
\bottomrule
\end{tabular}}
\label{tab:api}
\end{table}

To assess \ours's applicability with other image generation APIs, we additionally use SD+IPA and OJ (online) on COVIDx and KVASIR-f. Generative models behind different APIs are pre-trained on diverse large-scale public datasets~\cite{schuhmann2022laion, kakaobrain2022coyo}, resulting in varying knowledge. Despite these differences, \ours adapts to different generative APIs while maintaining its superiority, as shown in \cref{tab:api}. 
Since SD, SD+IPA, and OJ (online) share a similar generative backbone, with OJ (online) fine-tuned on SD, we infer that SD+IPA and OJ (online) have been exposed to more public data than SD, leading to broader knowledge. 
From \cref{tab:dataset} and \cref{tab:api}, we observe an interesting trend: \ours benefits from APIs trained on larger public datasets, whereas PE experiences a slight performance decline under the same conditions. 

\textbf{CAS \wrt Various Downstream Models}

\begin{figure}[h]
\centering
\includegraphics[width=\linewidth]{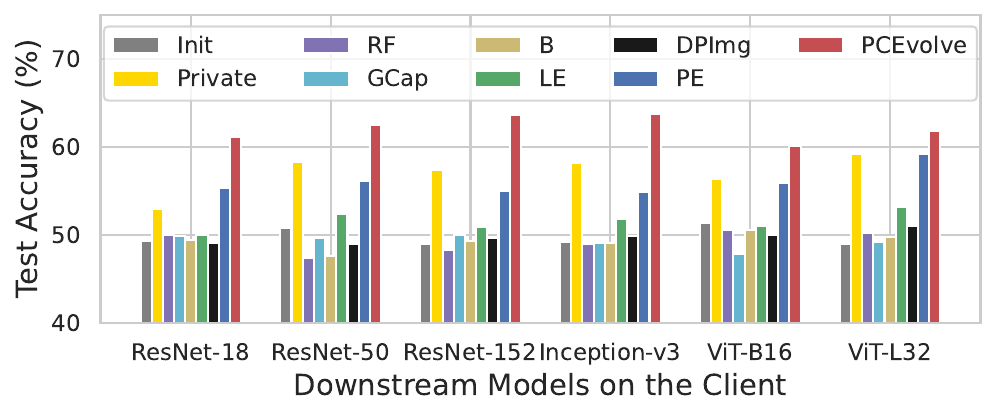}
\caption{Top-1 accuracy of various downstream models on COVIDx. ``Private'' represents an additional private baseline, which directly trains downstream models on few-shot private data. }
\label{fig:models}
\end{figure}

To evaluate the versatility of the synthetic dataset for downstream models, we consider six models that are widely used in specialized image domains~\cite{sarwinda2021deep, manzari2023medvit}, as shown in \cref{fig:models}. Following our setup in \cref{sec:setup}, we use pre-trained backbones and train newly initialized classification heads on the synthetic dataset. 
Here, we use the CLIP image encoder~\cite{radford2021learning} as the encoder $E_f$ following~\cite{lindifferentially}. 

In \cref{fig:models}, the synthetic image dataset generated by our \ours is compatible with various downstream models of different architectures (\eg, CNNs~\cite{lecun2015deep} and Transformers~\cite{vaswani2017attention}). Models trained on our \ours's synthetic dataset achieve the best performance among all counterparts. Notably, \ours surpasses the ``Private'' baseline by up to 8.20\%, which directly trains downstream models on few-shot private data for the same number of training steps as \ours. This demonstrates that our synthetic dataset incorporates valuable information from the generative API beyond what private data alone provides. In contrast, PE underperforms compared to ``Private'' when using larger models other than ResNet-18. While both \ours and PE improve downstream model performance over their initial states, most one-time generation methods degrade, except for LE. This highlights the substantial domain gap between synthetic and private datasets when an evolution loop is not used to incorporate sufficient private information. 

\subsection{Properties of \ours}
\label{sec:proper}

\begin{figure}[ht]
	\centering
	\hfill
	\subfigure[Initial]{\includegraphics[width=0.24\linewidth]{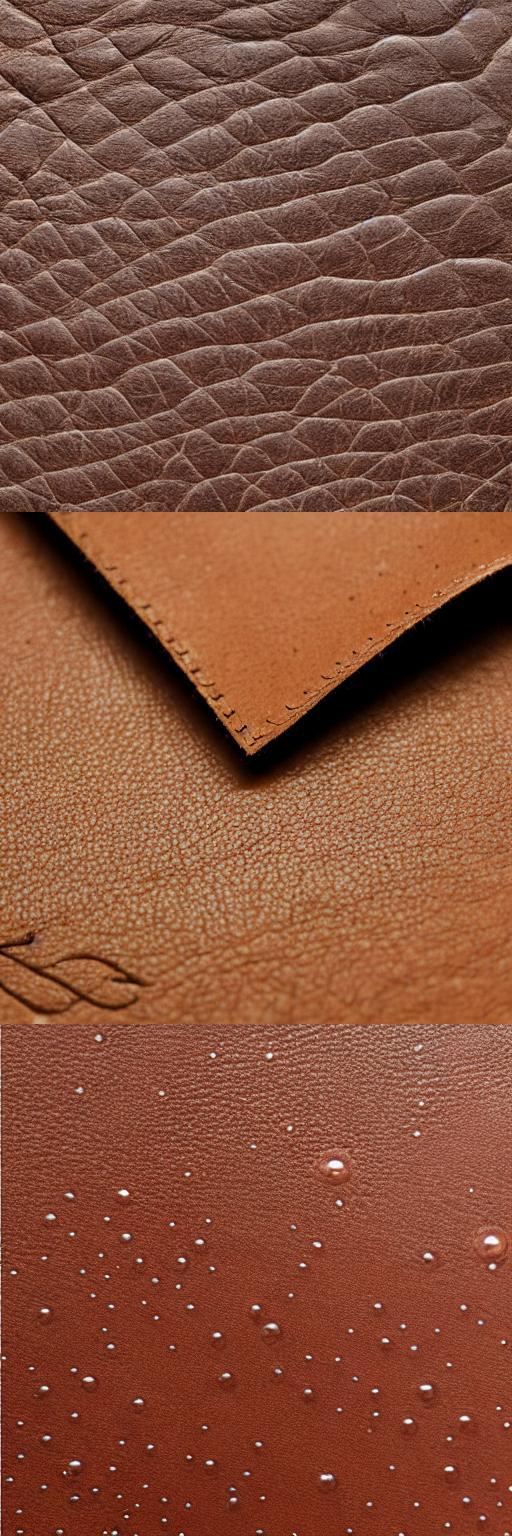}\label{fig:img-init}}
    \hfill
	\subfigure[PE]{\includegraphics[width=0.24\linewidth]{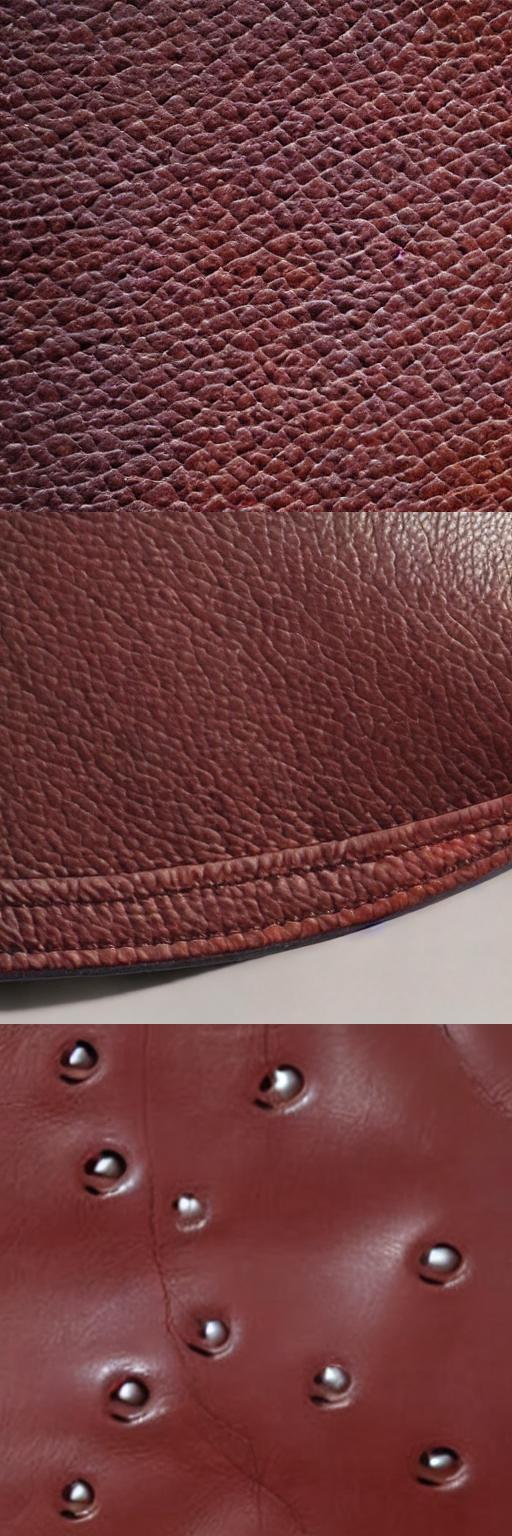}\label{fig:img-pe}}
	\hfill
    \hfill
	\subfigure[\ours]{\includegraphics[width=0.24\linewidth]{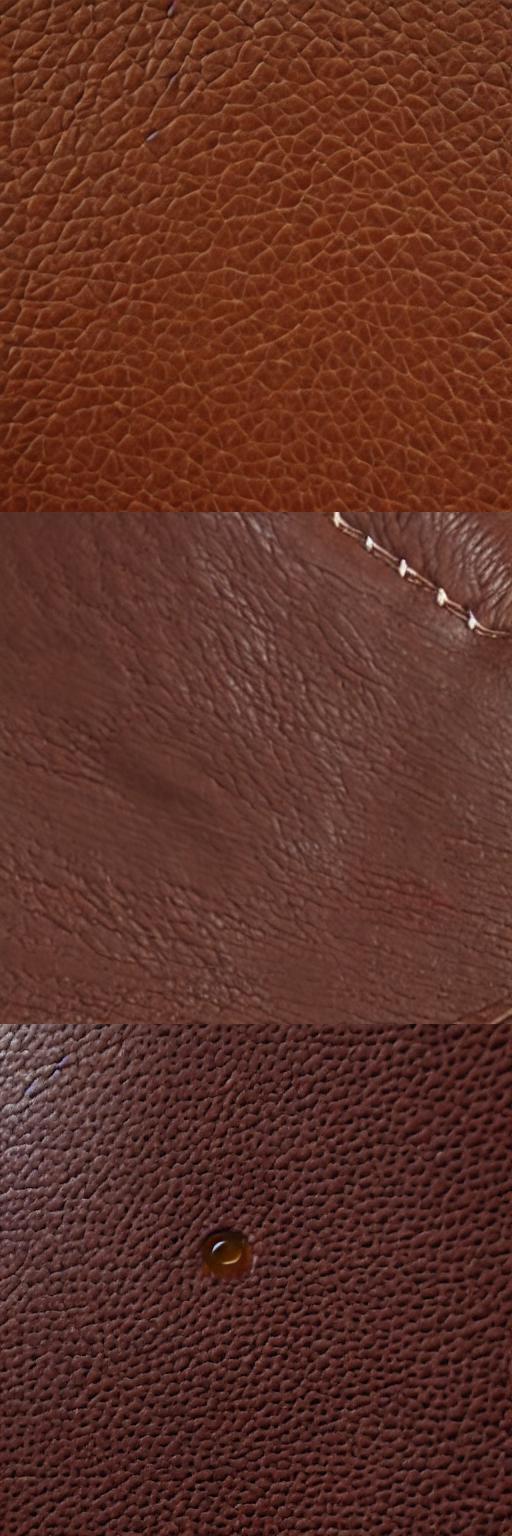}\label{fig:img-ours}}
    \hfill
	\subfigure[Private]{\includegraphics[width=0.24\linewidth]{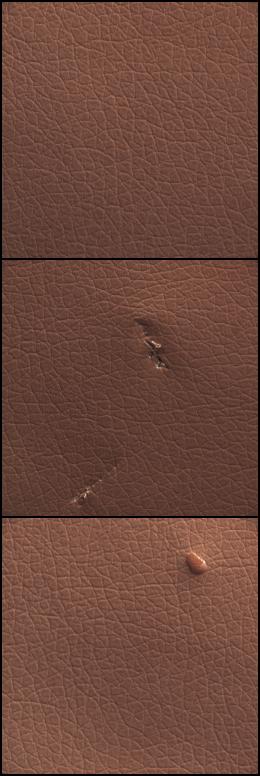}}
    \hfill
    \hfill
	\caption{Generated leather surface images \wrt MVAD-l for industry anomaly detection. The three rows show normal images, cut defects, and droplet defects. 
    ``Initial'' denotes the initial synthetic images in PE and \ours. 
    ``Private'' denotes the real images from MVAD-l.}
    \label{fig:img}
\end{figure}

\textbf{Synthetic Images}

As illustrated in \cref{fig:img}, the initial synthetic images differ significantly in meaning, content, color, and detail from few-shot private images. 
While PE employs an evolution loop to refine the initial images iteratively, its DP-protected similarity voting becomes nearly random with a few-shot private dataset. This limitation results in synthetic images that retain much information from the initial data while remaining distant from the private data. For instance, the synthetic images in \cref{fig:img-pe} fail to accurately depict the cut defect, instead emphasizing the boundary of a leather surface, as in \cref{fig:img-init}. Although the normal and droplet defect images align with the meaning of private images, they still differ in style and detail, leading to a significant domain gap. Such synthetic images fail to enhance downstream tasks and may even negatively impact them~\cite{hataya2023will}. 
In contrast, our \ours effectively tackles the few-shot challenge by leveraging inter-class contrastive relationships via the contrastive filter ($g$), resulting in class-discriminative leather images. As illustrated in \cref{fig:img-ours}, our synthetic images demonstrate greater diversity, such as varied lighting angles, across classes with additional knowledge extracted from APIs, while preserving a high degree of similarity to private images. This high similarity is achieved through our similarity calibrator ($h$), which enhances the likelihood of selecting the most similar candidates as prototypical synthetic images.

\textbf{Ablation Study} 

\begin{table}[h]
\centering
\caption{Top-1 accuracy (\%) on four datasets using \ours ($u = h\circ g$) variants with different $u$. $h'$ denotes the initial $h$ function in \cref{eq:h1}.}
\resizebox{\linewidth}{!}{
\begin{tabular}{l|c|ccc}
\toprule
 & $u = h\circ g$ & $u = g$ & $u = h'\circ g$ & $u = h$ \\
\midrule
COVIDx        & 64.04 & 56.59 & 56.11 & 55.58 \\
Came17     & 69.10 & 66.51 & 65.29 & 59.39 \\
KVASIR-f         & 50.95 & 44.61 & 50.67 & 47.78 \\
MVAD-l  & 59.26 & 55.74 & 55.56 & 53.71 \\
\bottomrule
\end{tabular}}
\label{tab:ab}
\end{table}

The design of the utility function $u$ is pivotal to \ours's effectiveness. In \cref{tab:ab}, we replace $u$ with alternative variants to demonstrate the significance of each component in \ours. Notably, the contrastive filter ($g$) plays an important role, achieving strong performance even when used alone. When combined with the similarity calibrator ($h$), \ours achieves up to a 7.45\% accuracy improvement on COVIDx. However, directly using the uncalibrated similarity scores from $h'\circ g$ leads to a performance drop of up to 7.93\% accuracy on COVIDx compared to \ours. Furthermore, removing $g$ leads to an even more significant performance drop, with accuracy decreasing by 9.71\% on Came17 at most. 

\textbf{Hyperparameter Study}

\begin{table}[h]
\centering
\caption{Top-1 accuracy (\%) on four datasets using \ours with varying $\tau$ values. }
\resizebox{\linewidth}{!}{
\begin{tabular}{l|ccccc}
\toprule
         & $\tau=1$ & $\tau=5$ & $\tau=10$ & $\tau=20$ & $\tau=100$ \\
\midrule
COVIDx          & 59.34    & 60.70    & 64.04     & 60.61     & 57.90      \\
Came17      & 65.85    & 66.84    & 69.10     & 68.42     & 68.05      \\
KVASIR-f        & 46.89    & 47.29    & 50.95     & 57.08     & 54.66      \\
MVAD-l  & 40.70    & 53.71    & 59.26     & 55.75     & 55.55      \\
\bottomrule
\end{tabular}}
\label{tab:hyper}
\end{table}

\ours has only one hyperparameter: the similarity calibrating factor $\tau$ in the similarity calibrator ($h$) (\cref{eq:h2}). 
As shown in \cref{tab:hyper}, selecting an appropriate $\tau$ enhances \ours's performance, with the optimal value being $\tau=10$ for COVIDx, Came17, and MVAD-l, while KVASIR-f achieves its best performance at $\tau=20$. 
When $\tau$ is too small (\eg, $\tau=1$), the $u$ scores (\ie, outputs of $h$) become similar among class-discriminative synthetic data, leading to nearly uniform selection probabilities after applying the EM $\mathcal{M}_u$ (\cref{def:exp}). Conversely, when $\tau$ is too large (\eg, $\tau=100$), most $u$ scores become zero, as $u$ computes $e^{-\ell\cdot \tau}$, where $\ell$ is the normalized $\ell_2$ distance in \cref{eq:h2}. In this case, only the best candidate (with $\ell = 0$) retains a $u$ score of $1$, while all others—including non-discriminative candidates—approach (near) $0$, inadvertently increasing the likelihood of selecting poor candidates according to \cref{def:exp}. To balance this trade-off, we choose an appropriate $\tau=10$ for \ours.

\section{Conclusion and Limitation}

Our proposed \ours effectively addresses the few-shot private data challenge in DP generation with APIs, particularly for specialized domains like healthcare and industry, as shown by \ours's superiority on four specialized datasets with various scenarios. By leveraging extra inter-class contrastive relationships in private data and proposing an adapted EM, \ours outperforms existing API-assisted methods, enabling high-quality DP synthetic images while leaving more practical few-shot scenarios for future exploration.


\section*{Acknowledgments}
This work was supported by the National Key R\&D Program of China under Grant No.2022ZD0160504, the Interdisciplinary Program of Shanghai Jiao Tong University (project number YG2024QNB05), and the Tsinghua University (AIR)-Asiainfo Technologies (China) Inc. Joint Research Center. 
We also thank Dr. Zinan Lin for his valuable support and guidance. 

\section*{Impact Statement}

This work highlights practical privacy-preserving generative API applications with a Differential Privacy (DP) guarantee, demonstrating the effectiveness of API-assisted training-free DP synthetic data generation in specialized domains. Apart from this contribution, we do not identify any significant societal implications that require specific attention.

\bibliography{main}

\begin{thebibliography}{82}
\providecommand{\natexlab}[1]{#1}
\providecommand{\url}[1]{\texttt{#1}}
\expandafter\ifx\csname urlstyle\endcsname\relax
  \providecommand{\doi}[1]{doi: #1}\else
  \providecommand{\doi}{doi: \begingroup \urlstyle{rm}\Url}\fi

\bibitem[Abou~Baker et~al.(2024)Abou~Baker, Rohrschneider, and Handmann]{abou2024parameter}
Abou~Baker, N., Rohrschneider, D., and Handmann, U.
\newblock Parameter-efficient fine-tuning of large pretrained models for instance segmentation tasks.
\newblock \emph{Machine Learning and Knowledge Extraction}, 6\penalty0 (4):\penalty0 2783--2807, 2024.

\bibitem[Barman et~al.(2024)Barman, Guo, and Conlan]{barman2024dark}
Barman, D., Guo, Z., and Conlan, O.
\newblock The dark side of language models: Exploring the potential of llms in multimedia disinformation generation and dissemination.
\newblock \emph{Machine Learning with Applications}, pp.\  100545, 2024.

\bibitem[Bergmann et~al.(2019)Bergmann, Fauser, Sattlegger, and Steger]{bergmann2019mvtec}
Bergmann, P., Fauser, M., Sattlegger, D., and Steger, C.
\newblock Mvtec ad--a comprehensive real-world dataset for unsupervised anomaly detection.
\newblock In \emph{IEEE Conference on Computer Vision and Pattern Recognition (CVPR)}, 2019.

\bibitem[Boland et~al.(2017)Boland, Karczewski, and Tatonetti]{boland2017ten}
Boland, M.~R., Karczewski, K.~J., and Tatonetti, N.~P.
\newblock Ten simple rules to enable multi-site collaborations through data sharing, 2017.

\bibitem[Brooks et~al.(2023)Brooks, Holynski, and Efros]{brooks2023instructpix2pix}
Brooks, T., Holynski, A., and Efros, A.~A.
\newblock Instructpix2pix: Learning to follow image editing instructions.
\newblock In \emph{IEEE Conference on Computer Vision and Pattern Recognition (CVPR)}, 2023.

\bibitem[Byeon et~al.(2022)Byeon, Park, Kim, Lee, Baek, and Kim]{kakaobrain2022coyo}
Byeon, M., Park, B., Kim, H., Lee, S., Baek, W., and Kim, S.
\newblock Coyo-700m: Image-text pair dataset.
\newblock \url{https://github.com/kakaobrain/coyo-dataset}, 2022.

\bibitem[Carlini et~al.(2022)Carlini, Chien, Nasr, Song, Terzis, and Tramer]{carlini2022membership}
Carlini, N., Chien, S., Nasr, M., Song, S., Terzis, A., and Tramer, F.
\newblock Membership inference attacks from first principles.
\newblock In \emph{2022 IEEE Symposium on Security and Privacy (SP)}, 2022.

\bibitem[Chen et~al.(2023)Chen, Tang, and Li]{chen2023industrial}
Chen, J., Tang, J., and Li, W.
\newblock Industrial edge intelligence: Federated-meta learning framework for few-shot fault diagnosis.
\newblock \emph{IEEE Transactions on Network Science and Engineering}, 10\penalty0 (6):\penalty0 3561--3573, 2023.

\bibitem[Chen et~al.(2022)Chen, Yu, Kao, Pang, and Lu]{chen2022dpgen}
Chen, J.-W., Yu, C.-M., Kao, C.-C., Pang, T.-W., and Lu, C.-S.
\newblock Dpgen: Differentially private generative energy-guided network for natural image synthesis.
\newblock In \emph{IEEE Conference on Computer Vision and Pattern Recognition (CVPR)}, 2022.

\bibitem[Chen et~al.(2020{\natexlab{a}})Chen, Kornblith, Norouzi, and Hinton]{chen2020simple}
Chen, T., Kornblith, S., Norouzi, M., and Hinton, G.
\newblock {A Simple Framework for Contrastive Learning of Visual Representations}.
\newblock In \emph{International Conference on Machine Learning (ICML)}, 2020{\natexlab{a}}.

\bibitem[Chen et~al.(2024)Chen, Tang, Zhu, Yan, Jin, Wang, Su, Zhang, Wang, and Tang]{chen2024janus}
Chen, X., Tang, S., Zhu, R., Yan, S., Jin, L., Wang, Z., Su, L., Zhang, Z., Wang, X., and Tang, H.
\newblock The janus interface: How fine-tuning in large language models amplifies the privacy risks.
\newblock In \emph{Proceedings of the 2024 on ACM SIGSAC Conference on Computer and Communications Security}, 2024.

\bibitem[Chen et~al.(2020{\natexlab{b}})Chen, Qin, Wang, Yu, and Gao]{chen2020fedhealth}
Chen, Y., Qin, X., Wang, J., Yu, C., and Gao, W.
\newblock Fedhealth: A federated transfer learning framework for wearable healthcare.
\newblock \emph{IEEE Intelligent Systems}, 35\penalty0 (4):\penalty0 83--93, 2020{\natexlab{b}}.

\bibitem[Croft et~al.(2021)Croft, Sack, and Shi]{croft2021obfuscation}
Croft, W.~L., Sack, J.-R., and Shi, W.
\newblock Obfuscation of images via differential privacy: From facial images to general images.
\newblock \emph{Peer-to-Peer Networking and Applications}, 14:\penalty0 1705--1733, 2021.

\bibitem[De~Cristofaro(2024)]{de2024synthetic}
De~Cristofaro, E.
\newblock Synthetic data: Methods, use cases, and risks.
\newblock \emph{IEEE Security \& Privacy}, 2024.

\bibitem[Dong et~al.(2020)Dong, Durfee, and Rogers]{dong2020optimal}
Dong, J., Durfee, D., and Rogers, R.
\newblock Optimal differential privacy composition for exponential mechanisms.
\newblock In \emph{International Conference on Machine Learning (ICML)}, 2020.

\bibitem[Duan et~al.(2024)Duan, Dziedzic, Papernot, and Boenisch]{duan2024flocks}
Duan, H., Dziedzic, A., Papernot, N., and Boenisch, F.
\newblock Flocks of stochastic parrots: Differentially private prompt learning for large language models.
\newblock \emph{Advances in Neural Information Processing Systems (NeurIPS)}, 2024.

\bibitem[Dwork(2008)]{dwork2008differential}
Dwork, C.
\newblock Differential privacy: A survey of results.
\newblock In \emph{International conference on theory and applications of models of computation}, 2008.

\bibitem[Dwork et~al.(2006)Dwork, McSherry, Nissim, and Smith]{dwork2006calibrating}
Dwork, C., McSherry, F., Nissim, K., and Smith, A.
\newblock Calibrating noise to sensitivity in private data analysis.
\newblock \emph{Theory of Cryptography}, pp.\  265--284, 2006.

\bibitem[Dwork et~al.(2014)Dwork, Roth, et~al.]{dwork2014algorithmic}
Dwork, C., Roth, A., et~al.
\newblock The algorithmic foundations of differential privacy.
\newblock \emph{Foundations and Trends{\textregistered} in Theoretical Computer Science}, 9\penalty0 (3--4):\penalty0 211--407, 2014.

\bibitem[Frolov et~al.(2021)Frolov, Hinz, Raue, Hees, and Dengel]{frolov2021adversarial}
Frolov, S., Hinz, T., Raue, F., Hees, J., and Dengel, A.
\newblock Adversarial text-to-image synthesis: A review.
\newblock \emph{Neural Networks}, 144:\penalty0 187--209, 2021.

\bibitem[Fu et~al.(2024{\natexlab{a}})Fu, Hong, Ling, Wang, Ran, Sun, Wang, Chen, and Cao]{fu2024differentially}
Fu, J., Hong, Y., Ling, X., Wang, L., Ran, X., Sun, Z., Wang, W.~H., Chen, Z., and Cao, Y.
\newblock Differentially private federated learning: A systematic review.
\newblock \emph{arXiv preprint arXiv:2405.08299}, 2024{\natexlab{a}}.

\bibitem[Fu et~al.(2024{\natexlab{b}})Fu, Ye, Hu, Chen, Wang, Wang, and Ran]{fu2024dpsur}
Fu, J., Ye, Q., Hu, H., Chen, Z., Wang, L., Wang, K., and Ran, X.
\newblock Dpsur: Accelerating differentially private stochastic gradient descent using selective update and release.
\newblock \emph{Proceedings of the VLDB Endowment}, 2024{\natexlab{b}}.

\bibitem[Gao et~al.(2023)Gao, Pi, Yong, Xu, Ye, Wu, ZHANG, Liang, Li, and Kong]{gaoself}
Gao, J., Pi, R., Yong, L., Xu, H., Ye, J., Wu, Z., ZHANG, W., Liang, X., Li, Z., and Kong, L.
\newblock Self-guided noise-free data generation for efficient zero-shot learning.
\newblock In \emph{International Conference on Learning Representations (ICLR)}, 2023.

\bibitem[Ghalebikesabi et~al.(2023)Ghalebikesabi, Berrada, Gowal, Ktena, Stanforth, Hayes, De, Smith, Wiles, and Balle]{ghalebikesabi2023differentially}
Ghalebikesabi, S., Berrada, L., Gowal, S., Ktena, I., Stanforth, R., Hayes, J., De, S., Smith, S.~L., Wiles, O., and Balle, B.
\newblock Differentially private diffusion models generate useful synthetic images.
\newblock \emph{arXiv preprint arXiv:2302.13861}, 2023.

\bibitem[Guan \& Liu(2021)Guan and Liu]{guan2021domain}
Guan, H. and Liu, M.
\newblock Domain adaptation for medical image analysis: a survey.
\newblock \emph{IEEE Transactions on Biomedical Engineering}, 69\penalty0 (3):\penalty0 1173--1185, 2021.

\bibitem[Hao et~al.(2024)Hao, Chi, Dong, and Wei]{hao2024optimizing}
Hao, Y., Chi, Z., Dong, L., and Wei, F.
\newblock Optimizing prompts for text-to-image generation.
\newblock \emph{Advances in Neural Information Processing Systems (NeurIPS)}, 2024.

\bibitem[Harmon et~al.(2020)Harmon, Sanford, Xu, Turkbey, Roth, Xu, Yang, Myronenko, Anderson, Amalou, et~al.]{harmon2020artificial}
Harmon, S.~A., Sanford, T.~H., Xu, S., Turkbey, E.~B., Roth, H., Xu, Z., Yang, D., Myronenko, A., Anderson, V., Amalou, A., et~al.
\newblock Artificial intelligence for the detection of covid-19 pneumonia on chest ct using multinational datasets.
\newblock \emph{Nature communications}, 11\penalty0 (1):\penalty0 4080, 2020.

\bibitem[Hataya et~al.(2023)Hataya, Bao, and Arai]{hataya2023will}
Hataya, R., Bao, H., and Arai, H.
\newblock Will large-scale generative models corrupt future datasets?
\newblock In \emph{IEEE International Conference on Computer Vision (ICCV)}, 2023.

\bibitem[He et~al.(2016)He, Zhang, Ren, and Sun]{he2016deep}
He, K., Zhang, X., Ren, S., and Sun, J.
\newblock {Deep Residual Learning for Image Recognition}.
\newblock In \emph{IEEE Conference on Computer Vision and Pattern Recognition (CVPR)}, 2016.

\bibitem[He et~al.(2023)He, Sun, Yu, Xue, Zhang, Torr, Bai, and Qi]{he2023synthetic}
He, R., Sun, S., Yu, X., Xue, C., Zhang, W., Torr, P., Bai, S., and Qi, X.
\newblock Is synthetic data from generative models ready for image recognition?
\newblock In \emph{International Conference on Learning Representations (ICLR)}, 2023.

\bibitem[Hou et~al.(2024)Hou, Shrivastava, Zhan, Conway, Le, Sagar, Fanti, and Lazar]{houpre}
Hou, C., Shrivastava, A., Zhan, H., Conway, R., Le, T., Sagar, A., Fanti, G., and Lazar, D.
\newblock Pre-text: Training language models on private federated data in the age of llms.
\newblock In \emph{International Conference on Machine Learning (ICML)}, 2024.

\bibitem[Hu et~al.(2020)Hu, Guo, Ratazzi, and Gong]{hu2020differentially}
Hu, R., Guo, Y., Ratazzi, E.~P., and Gong, Y.
\newblock Differentially private federated learning for resource-constrained internet of things.
\newblock \emph{arXiv preprint arXiv:2003.12705}, 2020.

\bibitem[Hu et~al.(2024)Hu, Wu, Li, Long, Garrido, Ge, Ding, Forsyth, Li, and Song]{hu2024sok}
Hu, Y., Wu, F., Li, Q., Long, Y., Garrido, G.~M., Ge, C., Ding, B., Forsyth, D., Li, B., and Song, D.
\newblock Sok: Privacy-preserving data synthesis.
\newblock In \emph{2024 IEEE Symposium on Security and Privacy (SP)}. IEEE, 2024.

\bibitem[Ji \& Chung(2024)Ji and Chung]{ji2024diffusion}
Ji, W. and Chung, A.~C.
\newblock Diffusion-based domain adaptation for medical image segmentation using stochastic step alignment.
\newblock In \emph{International Conference on Medical Image Computing and Computer-Assisted Intervention}. Springer, 2024.

\bibitem[Kather et~al.(2022)Kather, Ghaffari~Laleh, Foersch, and Truhn]{kather2022medical}
Kather, J.~N., Ghaffari~Laleh, N., Foersch, S., and Truhn, D.
\newblock Medical domain knowledge in domain-agnostic generative ai.
\newblock \emph{NPJ digital medicine}, 5\penalty0 (1):\penalty0 90, 2022.

\bibitem[Koh et~al.(2021)Koh, Sagawa, Marklund, Xie, Zhang, Balsubramani, Hu, Yasunaga, Phillips, Gao, et~al.]{koh2021wilds}
Koh, P.~W., Sagawa, S., Marklund, H., Xie, S.~M., Zhang, M., Balsubramani, A., Hu, W., Yasunaga, M., Phillips, R.~L., Gao, I., et~al.
\newblock Wilds: A benchmark of in-the-wild distribution shifts.
\newblock In \emph{International Conference on Machine Learning (ICML)}, 2021.

\bibitem[Kulis et~al.(2013)]{kulis2013metric}
Kulis, B. et~al.
\newblock Metric learning: A survey.
\newblock \emph{Foundations and Trends{\textregistered} in Machine Learning}, 5\penalty0 (4):\penalty0 287--364, 2013.

\bibitem[LeCun et~al.(2015)LeCun, Bengio, and Hinton]{lecun2015deep}
LeCun, Y., Bengio, Y., and Hinton, G.
\newblock {Deep Learning}.
\newblock \emph{Nature}, 521\penalty0 (7553):\penalty0 436--444, 2015.

\bibitem[Lee et~al.(2024)Lee, Yasunaga, Meng, Mai, Park, Gupta, Zhang, Narayanan, Teufel, Bellagente, et~al.]{lee2024holistic}
Lee, T., Yasunaga, M., Meng, C., Mai, Y., Park, J.~S., Gupta, A., Zhang, Y., Narayanan, D., Teufel, H., Bellagente, M., et~al.
\newblock Holistic evaluation of text-to-image models.
\newblock \emph{Advances in Neural Information Processing Systems (NeurIPS)}, 2024.

\bibitem[Li et~al.(2024)Li, Gong, Li, Zhao, Hou, and Wang]{li2024privimage}
Li, K., Gong, C., Li, Z., Zhao, Y., Hou, X., and Wang, T.
\newblock $\{$PrivImage$\}$: Differentially private synthetic image generation using diffusion models with $\{$Semantic-Aware$\}$ pretraining.
\newblock In \emph{33rd USENIX Security Symposium (USENIX Security 24)}, 2024.

\bibitem[Lin et~al.(2024)Lin, Gopi, Kulkarni, Nori, and Yekhanin]{lindifferentially}
Lin, Z., Gopi, S., Kulkarni, J., Nori, H., and Yekhanin, S.
\newblock Differentially private synthetic data via foundation model apis 1: Images.
\newblock In \emph{International Conference on Learning Representations (ICLR)}, 2024.

\bibitem[Liu et~al.(2023)Liu, Li, Wu, and Lee]{liu2024visual}
Liu, H., Li, C., Wu, Q., and Lee, Y.~J.
\newblock Visual instruction tuning.
\newblock \emph{Advances in Neural Information Processing Systems (NeurIPS)}, 2023.

\bibitem[Luo et~al.(2016)Luo, Chang, and Ban]{luo2016regression}
Luo, X., Chang, X., and Ban, X.
\newblock Regression and classification using extreme learning machine based on l1-norm and l2-norm.
\newblock \emph{Neurocomputing}, 174:\penalty0 179--186, 2016.

\bibitem[Manzari et~al.(2023)Manzari, Ahmadabadi, Kashiani, Shokouhi, and Ayatollahi]{manzari2023medvit}
Manzari, O.~N., Ahmadabadi, H., Kashiani, H., Shokouhi, S.~B., and Ayatollahi, A.
\newblock Medvit: a robust vision transformer for generalized medical image classification.
\newblock \emph{Computers in Biology and Medicine}, 157:\penalty0 106791, 2023.

\bibitem[McSherry \& Talwar(2007)McSherry and Talwar]{mcsherry2007mechanism}
McSherry, F. and Talwar, K.
\newblock Mechanism design via differential privacy.
\newblock In \emph{48th Annual IEEE Symposium on Foundations of Computer Science (FOCS)}, 2007.

\bibitem[Mesk{\'o}(2023)]{mesko2023prompt}
Mesk{\'o}, B.
\newblock Prompt engineering as an important emerging skill for medical professionals: tutorial.
\newblock \emph{Journal of medical Internet research}, 25:\penalty0 e50638, 2023.

\bibitem[Moor et~al.(2023)Moor, Banerjee, Abad, Krumholz, Leskovec, Topol, and Rajpurkar]{moor2023foundation}
Moor, M., Banerjee, O., Abad, Z. S.~H., Krumholz, H.~M., Leskovec, J., Topol, E.~J., and Rajpurkar, P.
\newblock Foundation models for generalist medical artificial intelligence.
\newblock \emph{Nature}, 616\penalty0 (7956):\penalty0 259--265, 2023.

\bibitem[OpenAI(2024)]{openai_finetuning_guide}
OpenAI.
\newblock Fine-tuning models - openai platform, 2024.
\newblock URL \url{https://platform.openai.com/docs/guides/fine-tuning#vision}.
\newblock Accessed: 2025-01-22.

\bibitem[Pan et~al.(2023)Pan, Zhou, and Tian]{pan2023arbitrary}
Pan, Z., Zhou, X., and Tian, H.
\newblock Arbitrary style guidance for enhanced diffusion-based text-to-image generation.
\newblock In \emph{Proceedings of the IEEE/CVF Winter Conference on Applications of Computer Vision}, 2023.

\bibitem[Pogorelov et~al.(2017)Pogorelov, Randel, Griwodz, Eskeland, de~Lange, Johansen, Spampinato, Dang-Nguyen, Lux, Schmidt, et~al.]{pogorelov2017kvasir}
Pogorelov, K., Randel, K.~R., Griwodz, C., Eskeland, S.~L., de~Lange, T., Johansen, D., Spampinato, C., Dang-Nguyen, D.-T., Lux, M., Schmidt, P.~T., et~al.
\newblock Kvasir: A multi-class image dataset for computer aided gastrointestinal disease detection.
\newblock In \emph{Proceedings of the 8th ACM on Multimedia Systems Conference}, 2017.

\bibitem[PromptHero(2023)]{openjourney}
PromptHero.
\newblock Openjourney, 2023.
\newblock URL \url{https://openjourney.art/}.

\bibitem[Qi et~al.(2024)Qi, Zeng, Xie, Chen, Jia, Mittal, and Henderson]{qifine}
Qi, X., Zeng, Y., Xie, T., Chen, P.-Y., Jia, R., Mittal, P., and Henderson, P.
\newblock Fine-tuning aligned language models compromises safety, even when users do not intend to!
\newblock In \emph{International Conference on Learning Representations (ICLR)}, 2024.

\bibitem[Radford et~al.(2021)Radford, Kim, Hallacy, Ramesh, Goh, Agarwal, Sastry, Askell, Mishkin, Clark, et~al.]{radford2021learning}
Radford, A., Kim, J.~W., Hallacy, C., Ramesh, A., Goh, G., Agarwal, S., Sastry, G., Askell, A., Mishkin, P., Clark, J., et~al.
\newblock Learning transferable visual models from natural language supervision.
\newblock In \emph{International Conference on Machine Learning (ICML)}, 2021.

\bibitem[Ravuri \& Vinyals(2019)Ravuri and Vinyals]{ravuri2019classification}
Ravuri, S. and Vinyals, O.
\newblock Classification accuracy score for conditional generative models.
\newblock \emph{Advances in Neural Information Processing Systems (NeurIPS)}, 2019.

\bibitem[Rombach et~al.(2022)Rombach, Blattmann, Lorenz, Esser, and Ommer]{rombach2022high}
Rombach, R., Blattmann, A., Lorenz, D., Esser, P., and Ommer, B.
\newblock High-resolution image synthesis with latent diffusion models.
\newblock In \emph{IEEE Conference on Computer Vision and Pattern Recognition (CVPR)}, 2022.

\bibitem[Ronanki et~al.(2024)Ronanki, Cabrero-Daniel, and Berger]{ronanki2024prompt}
Ronanki, K., Cabrero-Daniel, B., and Berger, C.
\newblock Prompt smells: An omen for undesirable generative ai outputs.
\newblock In \emph{Proceedings of the IEEE/ACM 3rd International Conference on AI Engineering-Software Engineering for AI}, 2024.

\bibitem[Roth et~al.(2022)Roth, Pemula, Zepeda, Sch{\"o}lkopf, Brox, and Gehler]{roth2022towards}
Roth, K., Pemula, L., Zepeda, J., Sch{\"o}lkopf, B., Brox, T., and Gehler, P.
\newblock Towards total recall in industrial anomaly detection.
\newblock In \emph{IEEE Conference on Computer Vision and Pattern Recognition (CVPR)}, 2022.

\bibitem[Samuel et~al.(2024)Samuel, Ben-Ari, Raviv, Darshan, and Chechik]{samuel2024generating}
Samuel, D., Ben-Ari, R., Raviv, S., Darshan, N., and Chechik, G.
\newblock Generating images of rare concepts using pre-trained diffusion models.
\newblock In \emph{AAAI Conference on Artificial Intelligence (AAAI)}, 2024.

\bibitem[Sander et~al.(2024)Sander, Yu, Sanjabi, Durmus, Ma, Chaudhuri, and Guo]{sanderdifferentially}
Sander, T., Yu, Y., Sanjabi, M., Durmus, A.~O., Ma, Y., Chaudhuri, K., and Guo, C.
\newblock Differentially private representation learning via image captioning.
\newblock In \emph{International Conference on Machine Learning (ICML)}, 2024.

\bibitem[Sarwinda et~al.(2021)Sarwinda, Paradisa, Bustamam, and Anggia]{sarwinda2021deep}
Sarwinda, D., Paradisa, R.~H., Bustamam, A., and Anggia, P.
\newblock Deep learning in image classification using residual network (resnet) variants for detection of colorectal cancer.
\newblock \emph{Procedia Computer Science}, 179:\penalty0 423--431, 2021.

\bibitem[Schuhmann et~al.(2022)Schuhmann, Beaumont, Vencu, Gordon, Wightman, Cherti, Coombes, Katta, Mullis, Wortsman, et~al.]{schuhmann2022laion}
Schuhmann, C., Beaumont, R., Vencu, R., Gordon, C., Wightman, R., Cherti, M., Coombes, T., Katta, A., Mullis, C., Wortsman, M., et~al.
\newblock Laion-5b: An open large-scale dataset for training next generation image-text models.
\newblock \emph{Advances in Neural Information Processing Systems (NeurIPS)}, 2022.

\bibitem[Seo et~al.(2024)Seo, Cho, Lee, Misra, Choi, Kim, and Choi]{seo2024just}
Seo, M., Cho, S., Lee, M., Misra, D., Choi, H., Kim, S.~J., and Choi, J.
\newblock Just say the name: Online continual learning with category names only via data generation.
\newblock \emph{arXiv preprint arXiv:2403.10853}, 2024.

\bibitem[Shokri et~al.(2017)Shokri, Stronati, Song, and Shmatikov]{shokri2017membership}
Shokri, R., Stronati, M., Song, C., and Shmatikov, V.
\newblock Membership inference attacks against machine learning models.
\newblock In \emph{2017 IEEE symposium on security and privacy (SP)}, 2017.

\bibitem[Song et~al.(2023)Song, Wang, Cai, Mondal, and Sahoo]{song2023comprehensive}
Song, Y., Wang, T., Cai, P., Mondal, S.~K., and Sahoo, J.~P.
\newblock A comprehensive survey of few-shot learning: Evolution, applications, challenges, and opportunities.
\newblock \emph{ACM Computing Surveys}, 55\penalty0 (13s):\penalty0 1--40, 2023.

\bibitem[Touvron et~al.(2023)Touvron, Martin, Stone, Albert, Almahairi, Babaei, Bashlykov, Batra, Bhargava, Bhosale, et~al.]{touvron2023llama}
Touvron, H., Martin, L., Stone, K., Albert, P., Almahairi, A., Babaei, Y., Bashlykov, N., Batra, S., Bhargava, P., Bhosale, S., et~al.
\newblock Llama 2: Open foundation and fine-tuned chat models.
\newblock \emph{arXiv preprint arXiv:2307.09288}, 2023.

\bibitem[Vaswani et~al.(2017)Vaswani, Shazeer, Parmar, Uszkoreit, Jones, Gomez, Kaiser, and Polosukhin]{vaswani2017attention}
Vaswani, A., Shazeer, N., Parmar, N., Uszkoreit, J., Jones, L., Gomez, A.~N., Kaiser, {\L}., and Polosukhin, I.
\newblock {Attention is All You Need}.
\newblock In \emph{Advances in Neural Information Processing Systems (NeurIPS)}, 2017.

\bibitem[Wang et~al.(2020)Wang, Lin, and Wong]{wang2020covid}
Wang, L., Lin, Z.~Q., and Wong, A.
\newblock Covid-net: A tailored deep convolutional neural network design for detection of covid-19 cases from chest x-ray images.
\newblock \emph{Scientific reports}, 10\penalty0 (1):\penalty0 19549, 2020.

\bibitem[Wang et~al.(2023)Wang, Shen, and Lim]{wang2023reprompt}
Wang, Y., Shen, S., and Lim, B.~Y.
\newblock Reprompt: Automatic prompt editing to refine ai-generative art towards precise expressions.
\newblock In \emph{Proceedings of the 2023 CHI conference on human factors in computing systems}, 2023.

\bibitem[Wornow et~al.(2023)Wornow, Xu, Thapa, Patel, Steinberg, Fleming, Pfeffer, Fries, and Shah]{wornow2023shaky}
Wornow, M., Xu, Y., Thapa, R., Patel, B., Steinberg, E., Fleming, S., Pfeffer, M.~A., Fries, J., and Shah, N.~H.
\newblock The shaky foundations of large language models and foundation models for electronic health records.
\newblock \emph{npj Digital Medicine}, 6\penalty0 (1):\penalty0 135, 2023.

\bibitem[Xie et~al.(2024)Xie, Lin, Backurs, Gopi, Yu, Inan, Nori, Jiang, Zhang, Lee, et~al.]{xiedifferentially}
Xie, C., Lin, Z., Backurs, A., Gopi, S., Yu, D., Inan, H.~A., Nori, H., Jiang, H., Zhang, H., Lee, Y.~T., et~al.
\newblock Differentially private synthetic data via foundation model apis 2: Text.
\newblock In \emph{International Conference on Machine Learning (ICML)}, 2024.

\bibitem[Xu et~al.(2022)Xu, Luo, Pan, Li, Pei, and Xu]{xu2022alleviating}
Xu, J., Luo, X., Pan, X., Li, Y., Pei, W., and Xu, Z.
\newblock Alleviating the sample selection bias in few-shot learning by removing projection to the centroid.
\newblock \emph{Advances in Neural Information Processing Systems (NeurIPS)}, 2022.

\bibitem[Yang et~al.(2023)Yang, Zhang, Song, Hong, Xu, Zhao, Zhang, Cui, and Yang]{yang2023diffusion}
Yang, L., Zhang, Z., Song, Y., Hong, S., Xu, R., Zhao, Y., Zhang, W., Cui, B., and Yang, M.-H.
\newblock Diffusion models: A comprehensive survey of methods and applications.
\newblock \emph{ACM Computing Surveys}, 56\penalty0 (4):\penalty0 1--39, 2023.

\bibitem[Yang et~al.(2021)Yang, Liu, and Xu]{yangfree}
Yang, S., Liu, L., and Xu, M.
\newblock Free lunch for few-shot learning: Distribution calibration.
\newblock In \emph{International Conference on Learning Representations (ICLR)}, 2021.

\bibitem[Ye et~al.(2023)Ye, Zhang, Liu, Han, and Yang]{ye2023ip}
Ye, H., Zhang, J., Liu, S., Han, X., and Yang, W.
\newblock Ip-adapter: Text compatible image prompt adapter for text-to-image diffusion models.
\newblock \emph{arXiv preprint arXiv:2308.06721}, 2023.

\bibitem[Ye et~al.(2022)Ye, Gao, Wu, Feng, Yu, and Kong]{ye2022progen}
Ye, J., Gao, J., Wu, Z., Feng, J., Yu, T., and Kong, L.
\newblock Progen: Progressive zero-shot dataset generation via in-context feedback.
\newblock In \emph{Findings of the Association for Computational Linguistics: EMNLP 2022}, 2022.

\bibitem[Zhang et~al.(2022)Zhang, Chen, and Li]{zhang2022privacy}
Zhang, J., Chen, Y., and Li, H.
\newblock Privacy leakage of adversarial training models in federated learning systems.
\newblock In \emph{IEEE Conference on Computer Vision and Pattern Recognition (CVPR)}, 2022.

\bibitem[Zhang et~al.(2023{\natexlab{a}})Zhang, Hua, Wang, Song, Xue, Ma, and Guan]{zhang2022fedala}
Zhang, J., Hua, Y., Wang, H., Song, T., Xue, Z., Ma, R., and Guan, H.
\newblock {FedALA: Adaptive Local Aggregation for Personalized Federated Learning}.
\newblock In \emph{AAAI Conference on Artificial Intelligence (AAAI)}, 2023{\natexlab{a}}.

\bibitem[Zhang et~al.(2024{\natexlab{a}})Zhang, Liu, Hua, and Cao]{zhang2024fedtgp}
Zhang, J., Liu, Y., Hua, Y., and Cao, J.
\newblock Fedtgp: Trainable global prototypes with adaptive-margin-enhanced contrastive learning for data and model heterogeneity in federated learning.
\newblock In \emph{AAAI Conference on Artificial Intelligence (AAAI)}, 2024{\natexlab{a}}.

\bibitem[Zhang et~al.(2025)Zhang, Liu, Hua, Wang, Song, Xue, Ma, and Cao]{zhang2025pfllib}
Zhang, J., Liu, Y., Hua, Y., Wang, H., Song, T., Xue, Z., Ma, R., and Cao, J.
\newblock Pfllib: A beginner-friendly and comprehensive personalized federated learning library and benchmark.
\newblock \emph{Journal of Machine Learning Research}, 26\penalty0 (50):\penalty0 1--10, 2025.

\bibitem[Zhang et~al.(2023{\natexlab{b}})Zhang, Rao, and Agrawala]{zhang2023adding}
Zhang, L., Rao, A., and Agrawala, M.
\newblock Adding conditional control to text-to-image diffusion models.
\newblock In \emph{IEEE International Conference on Computer Vision (ICCV)}, 2023{\natexlab{b}}.

\bibitem[Zhang et~al.(2024{\natexlab{b}})Zhang, Liang, Du, and Tian]{zhang2024dp}
Zhang, Y., Liang, X., Du, R., and Tian, J.
\newblock Dp-discriminator: A differential privacy evaluation tool based on gan.
\newblock In \emph{Proceedings of the 21st ACM International Conference on Computing Frontiers}, 2024{\natexlab{b}}.

\bibitem[Ziller et~al.(2021)Ziller, Usynin, Braren, Makowski, Rueckert, and Kaissis]{ziller2021medical}
Ziller, A., Usynin, D., Braren, R., Makowski, M., Rueckert, D., and Kaissis, G.
\newblock Medical imaging deep learning with differential privacy.
\newblock \emph{Scientific Reports}, 11\penalty0 (1):\penalty0 13524, 2021.

\end{thebibliography}
\bibliographystyle{icml2025}

\newpage
\appendix
\onecolumn





\section{Experimental Details}

We have included the necessary experimental details in the main body, and show more details here. 

\subsection{Image Generation APIs}
We consider three image generation APIs: Stable Diffusion (SD)~\cite{rombach2022high}, SD with the IP-Adapter (SD+IPA)~\cite{ye2023ip}, and the online OpenJourney (OJ) API~\cite{openjourney} (OJ (online)). Following PE~\cite{lindifferentially}, we manually implement SD on a server, providing an SD API with both text-to-image (t2i) and image-to-image (i2i) features. Specifically, we use the pre-trained open-source SD v1.5 model from HuggingFace\footnote{\url{https://huggingface.co/stable-diffusion-v1-5/stable-diffusion-v1-5}} and wrap it to expose only the API interface and serve as an API server, keeping the model details hidden from clients. SD+IPA is implemented similarly, by integrating a pre-trained open-source IP-Adapter\footnote{\url{https://huggingface.co/h94/IP-Adapter}} to the SD API. For the OJ (online) API, we use the getimg.ai platform to access the online OJ API\footnote{\url{https://dashboard.getimg.ai/models}}, which requires payments before usage. 
We retain the default settings (such as \linecode{num\_inference\_steps}=50, \linecode{guidance\_scale}=7.5, \etc) for all APIs to ensure better generalization. Similar to PE, we initialize the \linecode{strength} at 0.8 and anneal it to 0.6 with a 0.02 decrease per iteration for the i2i API. Additionally, we set \linecode{scale} to 0.5 for the IP-Adapter. Further details are available in our code. 

\subsection{Few-Shot Datasets}
We evaluate \ours on four datasets across two specialized domains under $K$-shot settings. In healthcare, we use (1) COVIDx\footnote{\url{https://www.kaggle.com/datasets/andyczhao/covidx-cxr2}}~\cite{wang2020covid} (chest X-ray images for COVID-19, two classes), (2) Camelyon17\footnote{\url{https://wilds.stanford.edu/datasets/\#camelyon17}}~\cite{koh2021wilds} (tumor tissue patches from breast cancer metastases, two classes), and (3) KVASIR-f\footnote{\url{https://datasets.simula.no/kvasir/}} (endoscopic images for gastrointestinal abnormal \underline{f}indings detection subset from KVASIR~\cite{pogorelov2017kvasir}, three classes). In industry, we use MVTecAD-l\footnote{\url{https://www.mvtec.com/company/research/datasets/mvtec-ad}} (\underline{l}eather surface anomaly detection subset from MVTecAD~\cite{bergmann2019mvtec}, three classes). Specifically, KVASIR-f is a subset of KVASIR containing all pathological \underline{f}inding images, while MVTecAD-l is a subset of MVTecAD focused on \underline{l}eather surface images. 
By default, we set $K=10$ for MVTecAD-l, containing only 19 leather surface images per class. We use 10 images for synthetic image generation and reserve the remaining 9 images for evaluating the test accuracy of the fine-tuned downstream models. This value of $K$ is typical for few-shot image tasks~\cite{he2023synthetic}. 
We create few-shot subsets from these datasets to represent realistic scenarios. In \cref{tab:dataset_info}, we list each dataset's details. We use a uniform simple text prompt $\mathcal{T} :=$ ``A DOMAIN image with LABEL'', where ``DOMAIN'' and ``LABEL'' are placeholders for respective domain and label names, for all datasets and tasks. More details are available in our code. 

\begin{table}[h]
\centering
\caption{The details of four few-shot datasets from two specialized domains. }
\resizebox{\linewidth}{!}{
\begin{tabular}{l|ccp{140pt}l}
\toprule
Dataset & Image Size & Test Set Size & Domain & Labels \\
\midrule
COVIDx & 256x256 & 8482 & ``chest radiography (X-ray)'' 
& [``'', ``COVID-19 pneumonia''] \\
Camelyon17 & 96x96 & 85054 & ``histological lymph node section'' & [``'', ``breast cancer with a tumor tissue''] \\
KVASIR-f & 256x256 & 600 & ``pathological damage in mucosa of gastrointestinal tract'' & [``esophagitis'', ``polyps'', ``ulcerative-colitis''] \\
MVTecAD-l & 256x256 & 27 & ``leather texture'' & [``'', ``cut defect'', ``droplet defect''] \\
\bottomrule
\end{tabular}}
\label{tab:dataset_info}
\end{table}

\subsection{Baselines}
We compare \ours with six baselines across three categories, all of which focus on generating image datasets using untrusted black-box API(s), without training: 

(\RomanNumeralCaps{1}) \textit{Using t2i APIs for image generation}: \\
$\bullet$~B~\cite{he2023synthetic}, which uses only a t2i API with a simple text prompt $\mathcal{T}$ that includes only the domain and class label name. 
$\bullet$~LE~\cite{seo2024just}, which extends B with a LLaMA~\cite{touvron2023llama} API to enhance $\mathcal{T}$. Specifically, we use an additional text prompt to enhance $\mathcal{T}$ with the LLaMA API: ``refine this description of images to introduce rich context: ''. 
$\bullet$~RF~\cite{samuel2024generating}, which filters out bad t2i-generated data that closely resemble private data from different classes. 
$\bullet$~GCap, which generates images using a t2i API with a LLaVA~\cite{liu2024visual} API for extracting private image captions. 

(\RomanNumeralCaps{2}) \textit{Using i2i APIs for image generation with DP}: \\
$\bullet$~DPImg, which directly adds DP (GM) noise to few-shot private images to generate DP replicas, which are then input to an i2i API. DPImg adapts RG~\cite{he2023synthetic} to ensure DP while avoiding modifications to the generative API. We compute the $\sigma$ for GM based on \cref{def:fD} given a total privacy cost $\epsilon_*$.

(\RomanNumeralCaps{3}) \textit{Using t2i and i2i APIs for image generation with DP}: \\
$\bullet$~PE~\cite{lindifferentially}, like \ours, generates DP synthetic image datasets using private data along with both t2i and i2i APIs within a privacy-preserving evolution loop. 
In few-shot scenarios, there are too few votes but too much noise in PE, making the thresholding operation on similarity votes meaningless. Therefore, we set $H=0$ for PE. Additionally, as shown in the PE paper, PE performs similarly for $H \geq 0$ when $\epsilon_* > 2$.

\subsection{Implementation Details}
To maximize performance in few-shot scenarios, following~\cite{he2023synthetic}, we train a new classification head on a pre-trained backbone model using the final synthetic dataset $\mathcal{D}_s$. The downstream model training during evaluation is done with a batch size of 16, a learning rate of 0.001, and 100 epochs. 
We report the Top-1 accuracy on the entire downstream test sets (see \cref{tab:dataset_info}). Top-1 accuracy, also known as the classification accuracy score (CAS)~\cite{ravuri2019classification}, is a widely used metric for assessing the quality of synthetic datasets in downstream tasks~\cite{frolov2021adversarial, lee2024holistic}. By default, we use ResNet-18~\cite{he2016deep, zhang2025pfllib, zhang2022fedala, zhang2024fedtgp} as the pre-trained backbone model and encoder $E_f$ due to its broad applicability across resource-constrained clients. 
For DP methods, the overall privacy cost $\epsilon_*$ is set to 10, 8, 8, and 10 for COVIDx, Camelyon17, KVASIR-f, and MVTecAD-l, respectively. For GM in DPImg and PE, we set $\delta$ to $10^{-5}$, higher than EM's $0$ in \ours. By default, we set the total generation iteration $T=20$ for \ours. 
Most of our experiments are run on a machine with 64 Intel(R) Xeon(R) Platinum 8362 CPUs, 256GB of memory, eight NVIDIA 3090 GPUs, and Ubuntu 20.04.4 LTS. While most experiments are completed within 48 hours, those involving a large $N$ for $N$-shot image generation may take up to a week. 

\section{CAS \wrt Overall Privacy Cost $\epsilon_*$}

\begin{table}[h]
\centering
\caption{Top-1 accuracy (\%) on four specialized datasets with varying overall privacy cost $\epsilon_*$. }
\resizebox{!}{!}{
\begin{tabular}{l|c|cccc}
\toprule
Dataset & DP Algorithm & $\epsilon_* = 4$ & $\epsilon_* = 8$ & $\epsilon_* = 10$ & $\epsilon_* = 20$ \\
\midrule
\multirow{2}{*}{COVIDx} & PE & 52.76 & 57.83 & 59.63 & 62.44 \\ 
 & \ours & 56.74 & 60.21 & 64.04 & 63.82 \\
\midrule
\multirow{2}{*}{Camelyon17} & PE & 62.29 & 63.66 & 63.37 & 65.29 \\ 
 & \ours & 68.05 & 69.11 & 69.58 & 69.95 \\ 
\midrule
\multirow{2}{*}{KVASIR-f} & PE & 43.72 & 48.88 & 51.01 & 51.83 \\ 
 & \ours & 50.44 & 50.95 & 51.67 & 52.11 \\ 
\midrule
\multirow{2}{*}{MVTecAD-l} & PE & 50.21 & 55.85 & 57.41 & 58.02 \\
 & \ours & 51.84 & 57.41 & 59.26 & 60.67 \\ 
\bottomrule
\end{tabular}}
\label{tab:eps}
\end{table}

To study the impact of the overall privacy cost $\epsilon_*$ on iterative generation algorithms, such as PE and our \ours, we vary $\epsilon_*$ and present the results in \cref{tab:eps}. We observe that both PE and \ours achieve lower accuracy with a smaller $\epsilon_*$ and perform better with larger values, consistent with the DP literature~\cite{dwork2014algorithmic}. To balance the privacy-utility trade-off, selecting an appropriate $\epsilon_*$ is crucial for different tasks and environments~\cite{lindifferentially}. 

\section{Iterative Improvement in the Evolution Loop}

\begin{figure*}[ht]
	\centering
	\subfigure[COVIDx]{\includegraphics[width=0.48\linewidth]{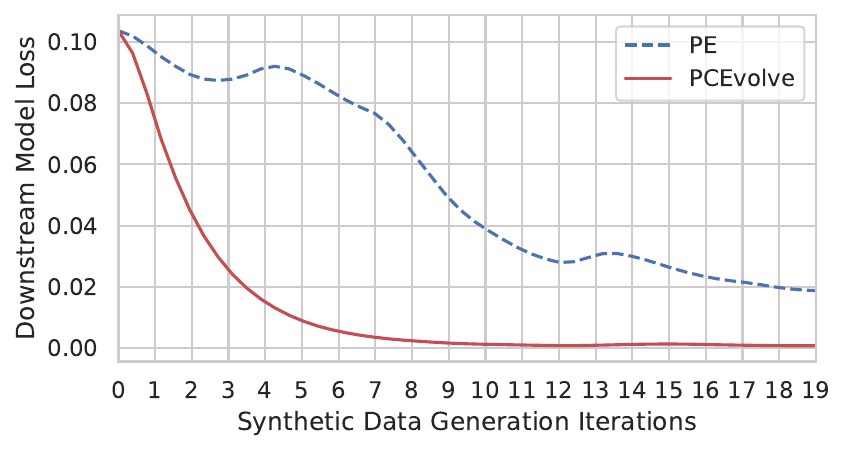}\label{fig:loss-covidx}}
	\subfigure[KVASIR-f]{\includegraphics[width=0.48\linewidth]{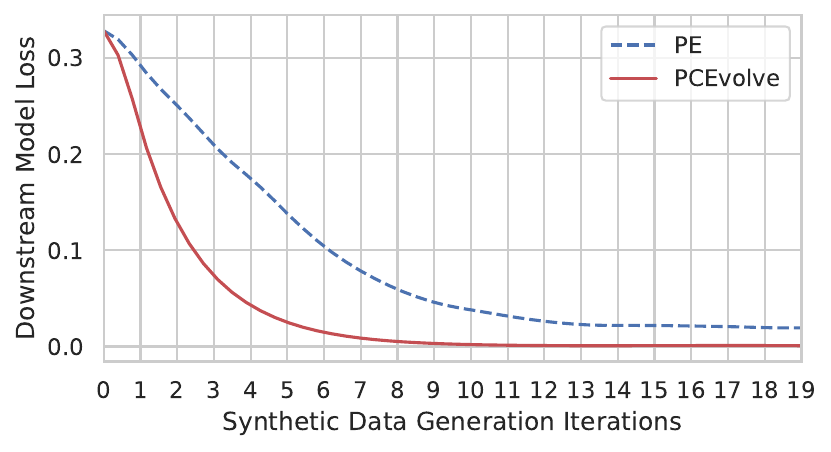}\label{fig:loss-kvasir}}
	\caption{The loss curves of ResNet-18~\cite{he2016deep}, which is \textit{retrained} at each iteration of synthetic data generation for algorithm performance evaluation. We use a CLIP image encoder~\cite{radford2021learning} as the encoder.}
    \label{fig:loss}
\end{figure*}

In the main body, following~\cite{he2023synthetic}, we train a new classification head for a pre-trained downstream model (\eg, ResNet-18) on the final synthetic dataset $\mathcal{D}_s$. To demonstrate the iterative improvement in the evolution loop, we retrain a new classification head for the given downstream model at each generation iteration and track the loss value after training on the synthetic data. The loss curves are shown in \cref{fig:loss}. Although the downstream model starts with the same initial loss value at the 0th iteration for all methods, our \ours rapidly reduces the loss in early iterations and consistently maintains a near-zero loss value in subsequent iterations, demonstrating steady iterative improvement. In contrast, PE also starts with the same initial loss but experiences fluctuations throughout the evolution process, ultimately reaching a higher loss. This is due to its GM-based similarity voting approach, which results in noisy synthetic images that hinder its performance. 

\section{CAS \wrt Various Encoders $E_f$}

\begin{figure}[h]
\centering
\includegraphics[width=0.5\linewidth]{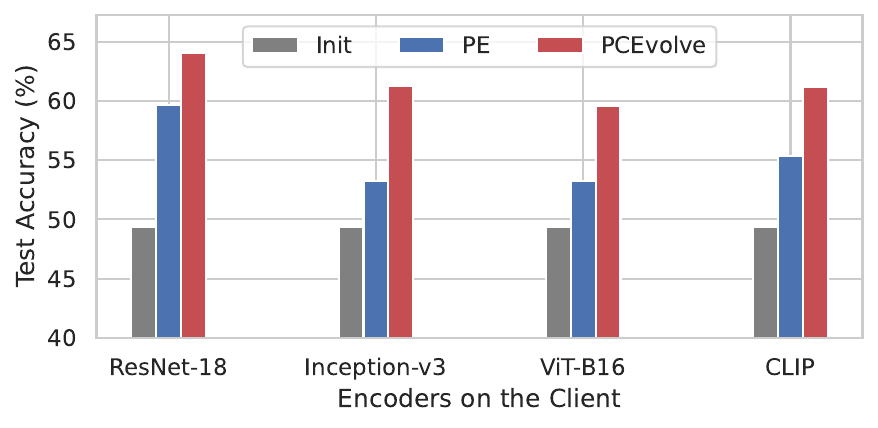}
\caption{Top-1 accuracy of ResNet-18 on COVIDx using four encoders. ``CLIP'' is short for CLIP image encoder. }
\label{fig:encoder}
\end{figure}

Here, we analyze PE and \ours to assess the impact of different encoders, which map images into feature vectors for efficient distance computation when evaluating synthetic data quality against private data. In \cref{fig:encoder}, we observe that the downstream model (ResNet-18) achieves optimal performance when paired with the same encoder (ResNet-18). When using other encoders, our \ours shows less performance degradation (4.50\%) compared to PE (6.41\%). A small encoder like ResNet-18 (0.01B parameters) is more practical than a large one like CLIP image encoder (0.3B parameters), as encoding is performed on resource-constrained clients and is required throughout the iterative synthetic data generation process. Both \ours and PE improve the performance of the initial downstream models with different encoders. 

\section{CAS on the Mixture of Synthetic and Private Data}

We follow PE~\cite{lindifferentially} to evaluate the quality of the synthetic image data in the main body, considering scenarios where synthetic data is widely utilized in various downstream tasks outside the private client, and the original private dataset is typically not accessible. In specific cases where private data owners wish to augment their local private datasets with synthetic data, the private data can be accessed, and the synthetic and private data can be mixed for augmentation.

\begin{table}[ht]
\centering
\caption{Top-1 accuracy (\%) on the mixture of synthetic and private data based on KVASIR-f. ``Syn'' is short for synthetic. }
\resizebox{!}{!}{
\begin{tabular}{l|c|c}
\toprule
& Syn & Syn + Private \\
\midrule
\textcolor{gray_}{Init} & \multicolumn{2}{c}{\textcolor{gray_}{33.43}} \\
\textcolor{gray_}{Private} & \multicolumn{2}{c}{\textcolor{gray_}{83.17}} \\
\midrule
RF       & 34.66 & 81.61 (\textcolor{red_}{-1.56}) \\
GCap     & 32.66 & 84.78 (\textcolor{green_}{+1.61}) \\
B      & 32.57 & 81.01 (\textcolor{red_}{-2.16}) \\
LE   & 35.51 & 84.61 (\textcolor{green_}{+1.44}) \\
\midrule
DPImg    & 33.35 & 62.94 (\textcolor{red_}{-20.23}) \\
\midrule
PE       & 48.88 & 88.67 (\textcolor{green_}{+5.50}) \\
\ours     & \textbf{50.95} & \textbf{90.51 (\textcolor{green_}{+7.34})} \\
\bottomrule
\end{tabular}}
\label{tab:mix}
\end{table}

For this scenario, we follow~\citet{he2023synthetic} and apply the mix training with the default downstream model (ResNet-18). As shown in \cref{tab:mix}, most baselines improve the performance when mixing synthetic and private data, with \ours showing the highest improvement by enhancing the class-discriminability of synthetic data, which benefits classification tasks. Although some t2i baselines (\eg, GCap and LE) perform poorly when evaluated on synthetic data alone, they show positive performance when applied to mixed datasets, as they bring additional valuable knowledge from APIs to the private data. In contrast, DPImg can negatively impact private data, as its synthetic data contains significant noise. 

\end{document}